\documentclass[12pt]{article}
\pdfoutput=1
\usepackage{fullpage}
\usepackage[margin=1.8cm]{geometry}                
\usepackage{graphicx}
\usepackage{amssymb}
\usepackage{amsmath}
\usepackage{amsthm}
\usepackage{MnSymbol}
\usepackage{hyperref}
\usepackage{epstopdf}
\usepackage{tikz}

\def\e{{\rm e}}
\def\ii{{\rm i}}
\def\dd{{\rm d}}
\def\be{\begin{equation}}
\def\ee{\end{equation}}
\def\bea{\begin{eqnarray}}
\def\eea{\end{eqnarray}}

\usepackage[font=small, labelfont=bf, width=0.8\textwidth]{caption}

\title{Discrete holomorphicity and integrability in loop models
\\[.05cm] 
with open boundaries}
\author{Jan de Gier$^1$, Alexander Lee$^1$ and J\o rgen Rasmussen$^2$\bigskip \\
\parbox{0.6\textwidth}{\center
\small $^1$\textit{Department of Mathematics and Statistics, The University of Melbourne, VIC 3010, Australia}
\medskip\\
$^2$\textit{School of Mathematics and Physics, University of Queensland,
St Lucia, Brisbane, Queensland 4072, Australia}}\bigskip}
\small\date{\small\today}                                          

\begin{document}
\maketitle
\abstract{We consider boundary conditions compatible with discrete holomorphicity for the dilute $O(n)$ and $C_2^{(1)}$ loop models. In each model, for a general set of boundary plaquettes, multiple types of loops can appear. A generalisation of Smirnov's parafermionic observable is therefore required in order to maintain the discrete holomorphicity property in the bulk. We show that there exist natural boundary conditions for this observable which are consistent with integrability, that is to say that, by imposing certain boundary conditions, we obtain a set of linear equations whose solutions also satisfy the corresponding reflection equation. In both loop models, several new sets of integrable weights are found using this approach.}
\let\thefootnote\relax\footnotetext{Email: \texttt{jdgier@unimelb.edu.au,\, a.lee19@pgrad.unimelb.edu.au,\, j.rasmussen@uq.edu.au}}

\section{Introduction}

The mathematically rigorous understanding of two-dimensional lattice models at criticality and their scaling limits has seen some significant recent breakthroughs, see e.\,g.~\cite{Smirnova,Smirnovb,Duminil-CopinS}. An important tool in this undertaking is the discretely holomorphic parafermionic observable introduced by Smirnov~\cite{Smirnova}.  The premise for the existence of such observables is as follows: at the continuum level, when the lattice spacing vanishes, many lattice models are believed to be conformally invariant and admit holomorphic observables. One may therefore hope to find observables in the finite lattice models that satisfy a \textit{discrete} form of holomorphicity.

Discretely holomorphic observables have the property that their discrete contour integral around any closed path vanishes. A similar property has been discussed by Bernard and Felder in the context of non-local conserved currents in lattice quantum field theories~\cite{BernardF}. However, the true power of discretely holomorphic observables was shown by Smirnov in proving the conformal invariance of the Ising model~\cite{SmirnovFermions}. They have since been used by Duminil-Copin and Smirnov to obtain a rigorous proof of the connective contstant of self-avoiding walks on the honeycomb lattice~\cite{DuminilSmir}, a value predicted 30 years earlier by Nienhuis~\cite{Nienhuis}. Other parafermionic observables and connections with SLE are discussed in~\cite{RivaCardy,LottiniR,IkhlefR}.

There is a remarkable but still not fully understood connection between discrete holomorphicity and Yang-Baxter integrability: by choosing a contour of integration surrounding a single plaquette of the lattice, one readily obtains the integrable Boltzmann weights of the model. This connection with discrete holomorphicity has been established by Rajabpour and Cardy~\cite{RajabCardy} in the $Z_N$ model, and by Ikhlef and Cardy~\cite{IkhlefCardy, Cardy} in the Potts model, the dense and dilute $O(n)$ models and the $C_2^{(1)}$ loop model. For a discussion of discrete holomorphicity in the six- and eight-vertex models, see~\cite{Tanhayi-AhariR}. In the case of the $Z_N$ model, Alam and Batchelor~\cite{AlamBatch} have recently shown that considering a discrete contour integral around two or three adjacent plaquettes leads to the inversion and star-triangle equations, respectively. We remark here that the work~\cite{BernardF} of Bernard and Felder mentioned above discusses integrability in the context of quantum groups and the related R-matrices, and that Nienhuis~\cite{Nienhuis90,Nienhuis2} derived integrable weights for the $O(n)$ loop models from a criticality argument.

An obvious question is whether there exist boundary conditions for discretely holomorphic observables that give rise to \textit{integrable boundary weights}, i.e. solutions to the reflection equation~\cite{Sklyanin}, a boundary version of the Yang-Baxter equation. Systematic studies of such solutions for interaction-round-a-face models and loop models are found in~\cite{BPO96} and~\cite{PRZ06,PRR08}. 

In~\cite{Beatona}, Beaton et al used a variant of discrete holomorphicity at the boundary to rigorously prove the critical temperature of the adsorption transition of self-avoiding walks on the honeycomb lattice, a value that had been obtained by Batchelor and Yung~\cite{Batchelor} by solving the corresponding reflection equation. Following the same reasoning as in~\cite{Beatona}, Beaton has also considered~\cite{Beatonb} a rotated honeycomb lattice with a boundary. Ikhlef imposed~\cite{Ikhlef} certain boundary conditions on a discretely holomorphic observable in the $O(n)$ and $Z_N$ models. In the $O(n)$ case, he recovered the integrable diagonal weights, and in the $Z_N$ case he found new integrable weights. 

The aim of this paper is to further the understanding of the connection between boundary integrability and boundary conditions of discretely holomorphic observables in the case of more general \textit{non-diagonal} boundary weights, where loops attached to the boundary may acquire different weights compared to loops in the bulk, such as in~\cite{DJSa,DJSb}. In the process, we slightly generalise the observables defined in the $O(n)$ and $C_2^{(1)}$ loop models and show they do indeed satisfy natural boundary conditions compatible with integrability. These boundary conditions give rise to a set of linear equations whose solutions satisfy the corresponding reflection equations. In this way, we not only recover the known solutions to the reflection equations, but also obtain new ones. 

The structure of this paper is as follows. In Section~\ref{SecDilute}, we review the dilute $O(n)$ model and the definition of the associated discretely holomorphic parafermionic observable. In Section~\ref{Sec3}, we extend this definition to ensure that this observable remains discretely holomorphic in the presence of non-trivial boundaries. In Section~\ref{Sec4}, we introduce two boundary conditions that the observable can satisfy, and use these to obtain the integrable boundary weights of the model. In Section~\ref{Sec5}, we repeat this procedure for the $C_2^{(1)}$ loop model. Section~\ref{SecConcl} contains some concluding remarks, while a discussion of the reflection equations and a further generalisation of the dilute $O(n)$ model are deferred to Appendix~\ref{AppReflection}.

\section{The dilute \boldmath{$O(n)$} loop model}
\label{SecDilute}

\subsection{Description of the model}

\begin{figure}
\centering
\begin{picture}(200,100)
\put(0,25){\includegraphics[scale=0.35]{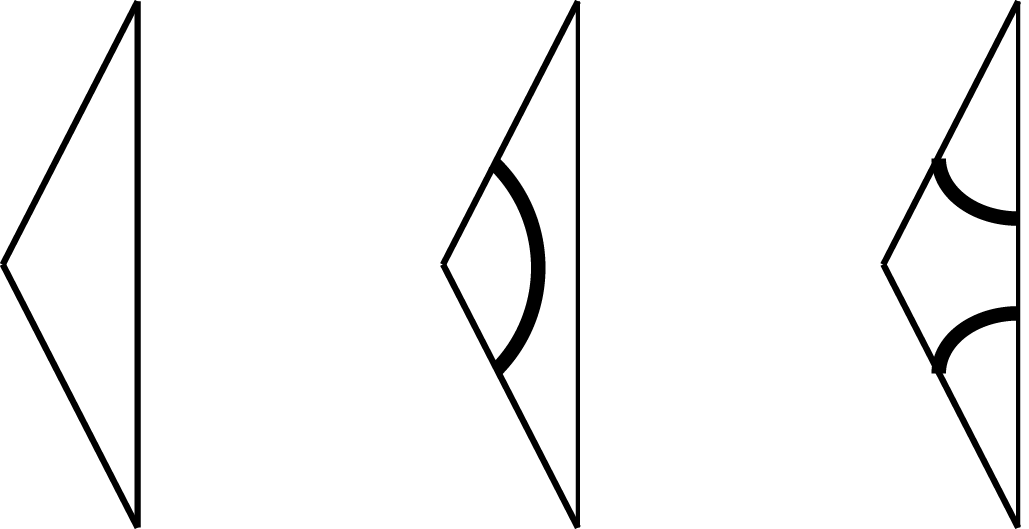}}
\put(15,0){$\beta_1$}
\put(90,0){$\beta_2$}
\put(163,0){$\beta_3$}
\end{picture}
\caption{The boundary plaquettes of the dilute $O(n)$ model along with the Boltzmann weights.}
\label{fig:diluteboundary}
\end{figure}
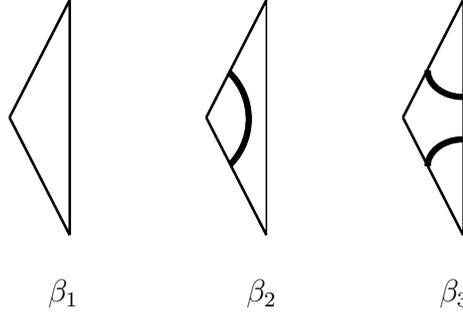

The dilute $O(n)$ model is a lattice model of closed non-intersecting, non-oriented loops. A given configuration of the model consists of a choice of tiling of the domain $\Omega$ (see below) by the bulk plaquettes shown in Fig.~\ref{fig:dilutebulk} and the boundary plaquettes shown in Fig.~\ref{fig:diluteboundary}. 
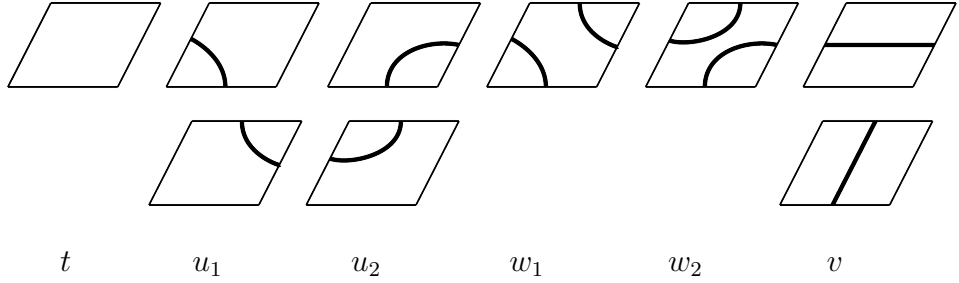
\begin{figure}
\centering
\begin{picture}(350,100)
\put(0,25){\includegraphics[scale=0.25]{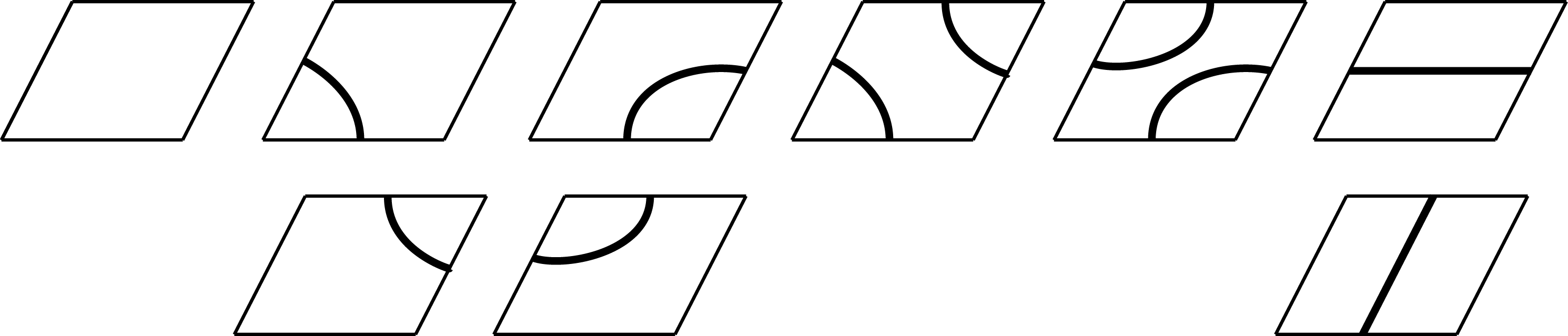}}
\put(20,0){$t$}
\put(70,0){$u_1$}
\put(130,0){$u_2$}
\put(190,0){$w_1$}
\put(250,0){$w_2$}
\put(310,0){$v$}
\end{picture}
\caption{The bulk plaquettes of the dilute $O(n)$ model, labelled by their corresponding Boltzmann weights.}
\label{fig:dilutebulk}
\end{figure}

Although more could be considered, we allow for five types of loops in the model:
\begin{itemize}
\item Bulk loops that lie entirely in the bulk. These loops carry the weight $n$. 
\item Bulk loops that pass through the boundary via $\beta_2$ plaquettes. These loops also carry the weight $n$.
\item Boundary loops that are attached to the boundary at a single $\beta_3$ boundary plaquette, or at two different $\beta_3$ plaquettes in such a way that the top part of the loop lies in the top half of the topmost of the two boundary plaquettes. These loops receive the weight $n_1$.
\item Boundary loops that are attached to the boundary at two different $\beta_3$ boundary plaquettes such that the top half of the loop lies in the bottom half of the topmost of the two boundary plaquettes. These loops receive the weight $n_2$. 
\item Boundary loops that are attached to two different $\beta_3$ plaquettes such that both the top and bottom of the boundary loop are attached to either the top edges of the $\beta_3$ plaquettes or the bottom edges. These loops receive the weight $n_3$. This type of loop occurs only if open loop segments are allowed and will therefore not play a role before we extend the model in Section~\ref{Sec3}.
\end{itemize}

\begin{figure}[ht]
\centering
\begin{picture}(100,200)
\put(0,0){\includegraphics[scale=0.25]{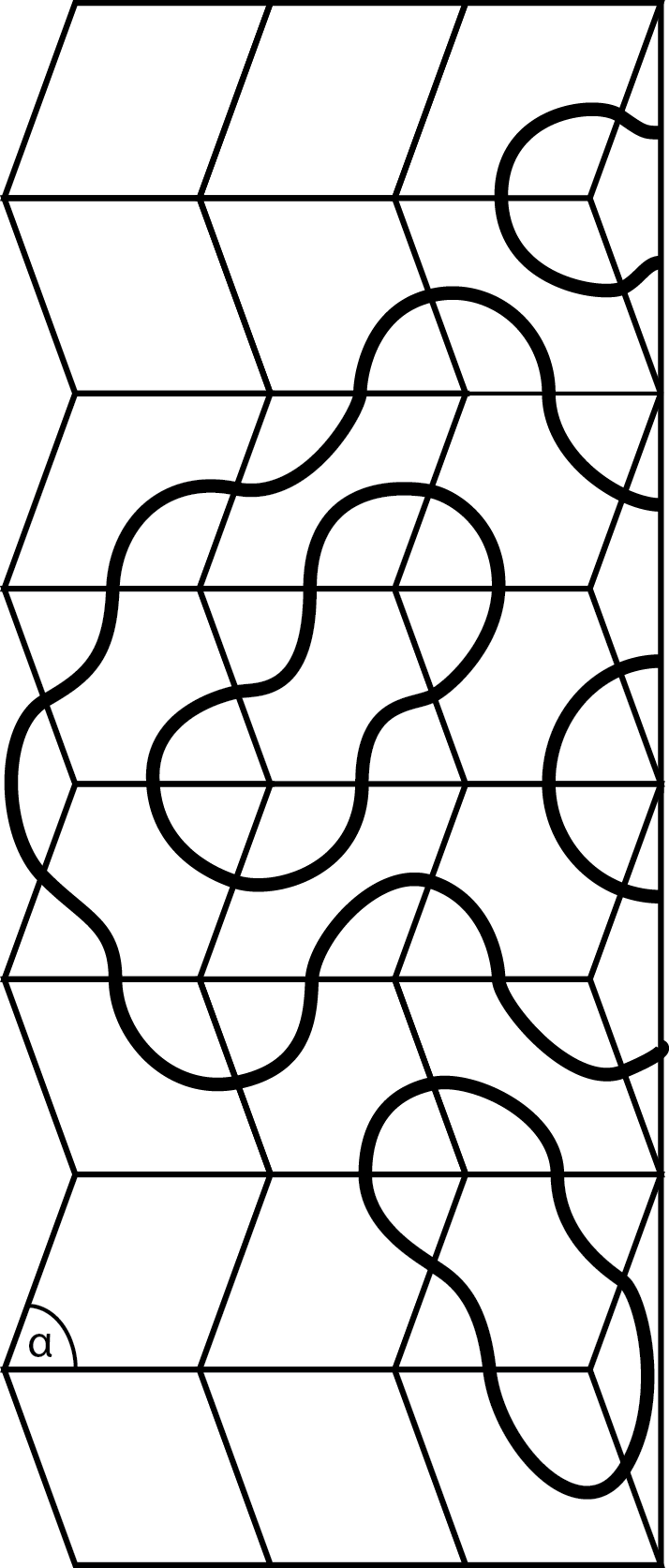}}
\end{picture}
\caption{A subdomain of the domain $\Omega$ with the four types of loops shown: a bulk loop, of weight $n$, passing through bulk plaquettes only; a bulk loop, of weight $n$, passing through a $\beta_2$ boundary plaquette; a boundary loop, of weight $n_2$, whose top endpoint lies in the bottom half of a $\beta_3$ boundary plaquette; and a pair of boundary loops, of weight $n_1$, whose top endpoints lie in top halves of $\beta_3$ boundary plaquettes. The acute angle of the rhombi is given by $\alpha$.}
\label{fig:domain}
\end{figure}
Let us also define the domain $\Omega$: starting from a regular square lattice, we skew adjacent rows in an opposite sense and scale the domain so all the faces become rhombi with acute angle $\alpha$. There is also a right-hand boundary consisting of isosceles triangles. A small portion of $\Omega$ is shown in Fig.~\ref{fig:domain}. For $\alpha=\frac{\pi}{2}$, the domain becomes the regular square lattice with a single straight-edged boundary. 

The total weight $P(G)$ of a configuration $G$ is the product of three contributions: 
\bea
 P(G)&=&\Big(t^{\#[t\mbox{{\scriptsize -plaquettes}}]}\ldots v^{\#[v\mbox{{\scriptsize -plaquettes}}]}\Big)
  \,\Big(\beta_1^{\#[\beta_1\mbox{{\scriptsize -plaquettes}}]}
   \beta_2^{\#[\beta_2\mbox{{\scriptsize -plaquettes}}]}
   \beta_3^{\#[\beta_3\mbox{{\scriptsize -plaquettes}}]}\Big)\nonumber\\[.1cm]
 &\times&\Big(n^{\#[n\mbox{{\scriptsize -loops}}]}
  n_1^{\#[n_1\mbox{{\scriptsize -loops}}]}n_2^{\#[n_2\mbox{{\scriptsize -loops}}]}\Big).
\label{PG}
\eea
The first contribution is the product of the \textit{local bulk} plaquette weights; the second contribution is the product of the \textit{local boundary} plaquette weights; while the third contribution is the product of the \textit{non-local} loop weights. Finally, the partition function is the sum over all possible configurations:
\be
 Z=\sum_{G}P(G).
\ee

Although we are working in a slightly different domain, the loop model with the boundary plaquettes just described was introduced by Dubail, Jacobsen and Saleur~\cite{DJSa,DJSb}. We generalise this loop model in Section~\ref{Sec3} and further in Appendix~\ref{AppAsymmOn}.
%

\subsection{Parafermionic observable}

For the dilute $O(n)$ model with only bulk loops, Smirnov introduced a discrete, complex-valued observable defined as follows. 

One first introduces a defect consisting of an open, oriented loop segment starting at a base point, here set to be $-\infty$. A configuration $\gamma$ is then a collection of closed loops in addition to this defect. The observable is defined at each of the mid-edges of the lattice by
\be
 F(z) := \sum_{\gamma:\,-\infty \to z}P(\gamma)\e^{-\ii s W(\gamma)},
\label{eq:Fdef}
\ee
where the sum is over configurations $\gamma$ for which the defect ends at $z$. $P(\gamma)$ is the Boltzmann weight of the configuration $\gamma$, while $W(\gamma)$ is the winding angle of the defect from the base point $-\infty$ to $z$. We define the initial winding angle to be $0$ along the positive real axis. The parameter $s$ is referred to as the spin, giving rise to the term $\emph{parafermionic}$ observable.

Smirnov introduced \eqref{eq:Fdef} with the aim of constructing a lattice observable that satisfies the discrete holomorphicity condition in the bulk:
\be
 \sum_{i}F(z_i)\Delta z_i=0,
\label{eqn:discholom}
\ee
where the sum runs over the edges of a plaquette or a cluster of plaquettes in a counterclockwise manner. Ikhlef and Cardy~\cite{IkhlefCardy} observed that \eqref{eqn:discholom} precisely holds for the \textit{integrable} $O(n)$ model, i.e. when the Boltzmann weights satisfy the Yang-Baxter equation. Here we recall their argument. 

\begin{figure}[ht]
\centering
\includegraphics[scale=0.25]{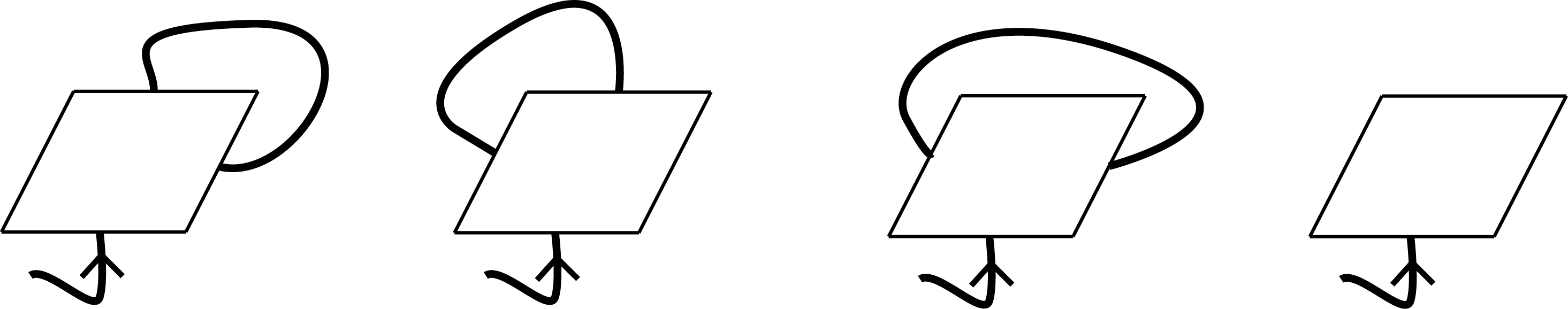}
\caption{The bulk plaquettes of the dilute $O(n)$ model: bulk connectivities.}
\label{fig:extcon1}
\end{figure}

\begin{figure}
\begin{center}
\includegraphics[scale=0.25]{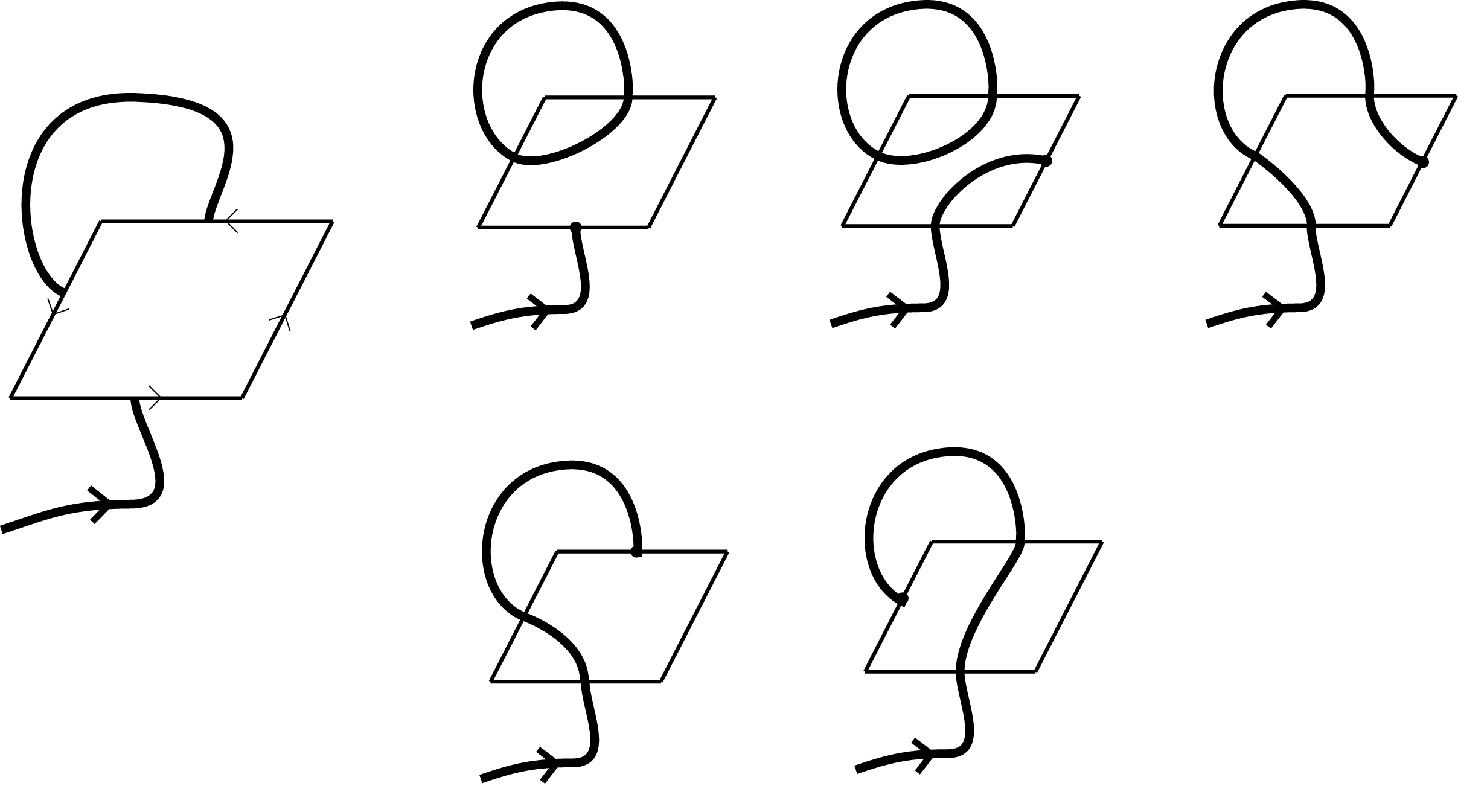}
\end{center}
\caption{The possible internal connectivities for the external configuration indicated by the second diagram in Fig.~\ref{fig:extcon1}, giving rise to the second equation in (\ref{eq:DHbulk1}).}
\label{fig:intconf}
\end{figure}

The configurations appearing in the bulk dilute $O(n)$ model arise from considering the four distinct ways a loop configuration external to a single plaquette can be fixed, see Fig.~\ref{fig:extcon1}. 
To illustrate, let us consider the case described in Fig.~\ref{fig:intconf}, and let $P(G^*)$ denote the total Boltzmann weight of the configuration \textit{external} to the plaquette. The discrete contour integral 
arising from the local configurations in Fig.~\ref{fig:intconf} is then given by
\begin{equation}
 \sum_{i}F(z_i)\Delta z_i=P(G^*)\, \e^{-s\ii\pi/2} \big[ nu_2 + \zeta\xi (n w_2 + w_1) - \xi u_1 -\zeta \overline{\xi} v\big].
\end{equation}
where 
\be
 \xi=\e^{\ii s \pi},\qquad \zeta=\e^{-\ii \alpha(s-1)}. 
\ee

{}From now on, we will discard overall irrelevant pre-factors such as $P(G^*)$. Thus, the following complex equations are obtained by requiring discrete holomorphicity (\ref{eqn:discholom}), where the discrete contour integral follows from going around a \textit{single} plaquette in a counterclockwise manner in each of the four cases of Fig.~\ref{fig:extcon1}:
\be
\begin{array}{rcl}
 nu_1 + \zeta\xi^2 v - \overline{\xi} u_2 -\zeta (w_2 + n w_1) &=& 0, \\[.1cm]
 nu_2 + \zeta\xi (n w_2 + w_1) - \xi u_1 -\zeta \overline{\xi} v &=& 0, \\[.1cm]
 nv + \zeta \xi^2 u_1 - (\xi^2 w_1 + \overline{\xi}^2 w_2) - \zeta\overline{\xi} u_2 &=& 0,\\[.1cm]
 t+\zeta\xi u_2 - v - \zeta u_1 &=& 0.
\end{array}
\label{eq:DHbulk1}
\ee
Assuming \textit{real} Boltzmann weights, and parameterising the loop fugacity $n$ by
\be
 n=-2\cos4\lambda,
\ee
it follows from simple linear algebra that \eqref{eq:DHbulk1} only has a non-trivial solution if $s=3\lambda/\pi+1$. 

For convenience, we now introduce the $\emph{spectral parameter}$
\be
x=\alpha(s-1), 
\ee
and write $\zeta=\e^{-\ii x}$. The Boltzmann weights that solve \eqref{eq:DHbulk1} precisely correspond to the well-known integrable weights

\be
\begin{array}{rcl}
 t(x) &=& \sin x \sin(3\lambda-x)+\sin2\lambda\sin3\lambda,\\[.1cm] 
 u_1(x) &=& \sin2\lambda\sin(3\lambda-x), \\[.1cm]
 u_2(x) &=& \sin2\lambda\sin x, \\[.1cm]
 v(x) &=& \sin x \sin(3\lambda-x), \\[.1cm]
 w_1(x) &=& \sin(2\lambda-x)\sin(3\lambda-x), \\[.1cm]
 w_2(x) &=& -\sin x \sin(\lambda-x),
\end{array}
\label{eq:BoltzWeightsBulk}
\ee
arising as the one-parameter family of solutions to the Yang-Baxter equation obtained in~\cite{Nienhuis90,Nienhuis2}.

\section{Discrete holomorphicity with a boundary}
\label{Sec3}

\subsection{Bulk configurations}

The introduction of the $\beta_3$ plaquette results in a large number of additional bulk configurations as a loop segment may now pass through the boundary via these new boundary plaquettes. We now also have to modify the definition of the parafermionic observable to account for this and still be able to require that the observable satisfies the discrete holomorphicity condition in the bulk (\ref{eqn:discholom}).

\begin{figure}[ht]
\centering
\includegraphics[scale=0.25]{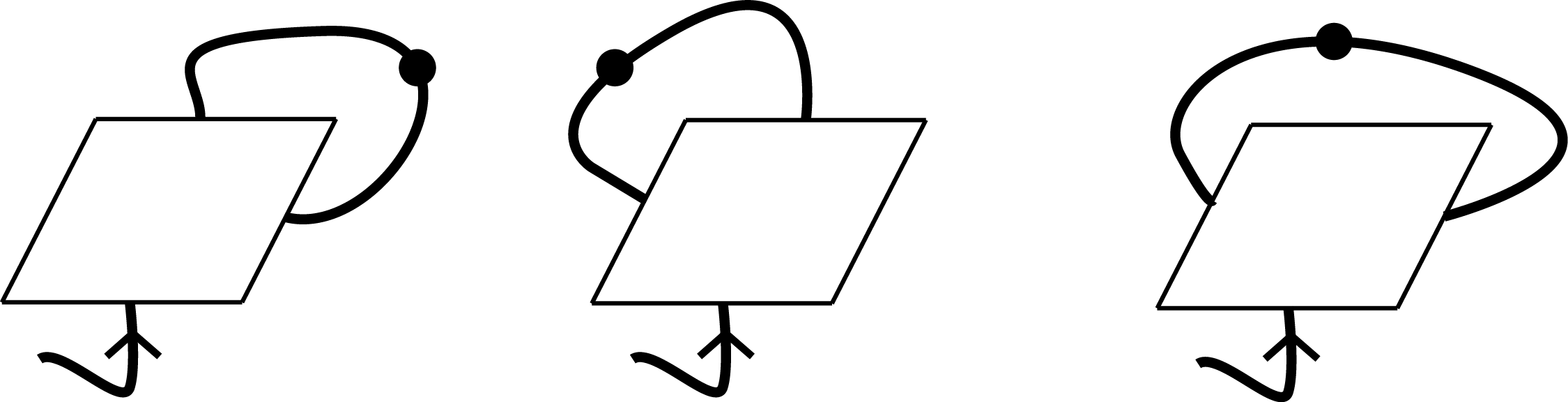}
\caption{The bulk plaquettes of the dilute $O(n)$ model: boundary connectivities.}
\label{fig:extcon2}
\end{figure}
In the following we refer to the bulk plaquette that we are taking the contour sum around as simply ``the bulk plaquette"$\!$. Firstly, consider the case where the external loop configuration is attached to the boundary via a single $\beta_3$ plaquette. In Fig.~\ref{fig:extcon2}, we indicate this by placing a blob on the loop segment.  In such cases, two things may happen. We may obtain a boundary loop weight $n_1$ by closing the loop in the interior of the bulk plaquette. Alternatively, if the defect becomes connected to the $\beta_3$ plaquette, we say that it ``passes through" the $\beta_3$ plaquette before re-entering the bulk plaquette. In such cases, we introduce an additional factor $q_1$ ($\overline{q}_1$) in the parafermionic observable according to whether the defect has passed counterclockwise (clockwise) through the $\beta_3$ plaquette, see (\ref{eq:Fdefb}) below.
\begin{figure}[ht]
\centering
\includegraphics[scale=0.25]{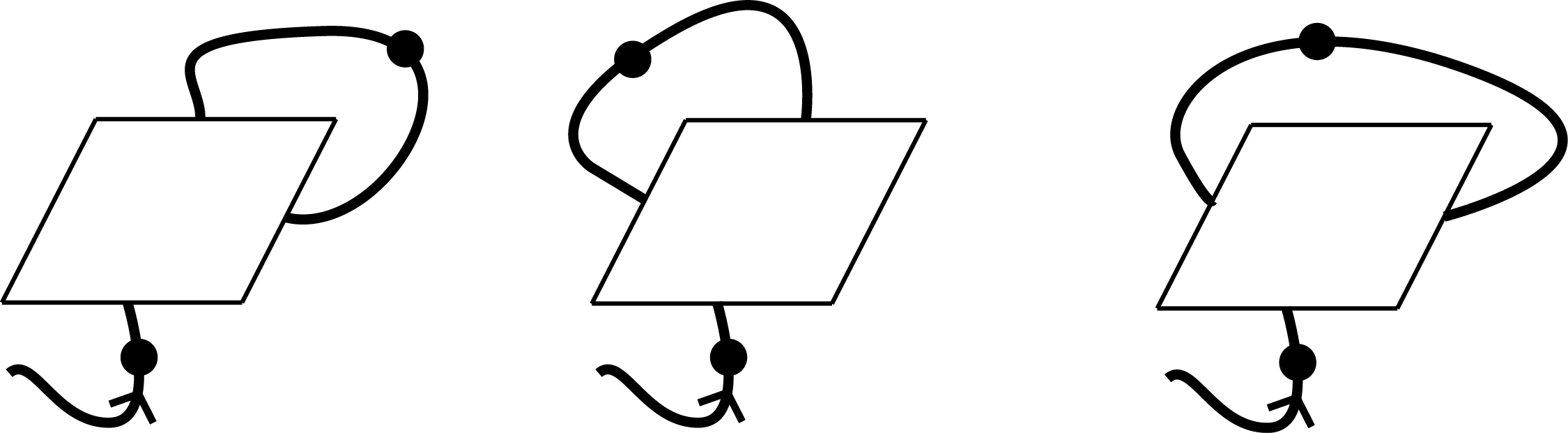}
\caption{Defect passing through a $\beta_3$ boundary plaquette before (and possibly again after) entering the bulk plaquette.}
\label{fig:extcon3}
\end{figure}

\begin{figure}[ht]
\centering
\includegraphics[scale=0.25]{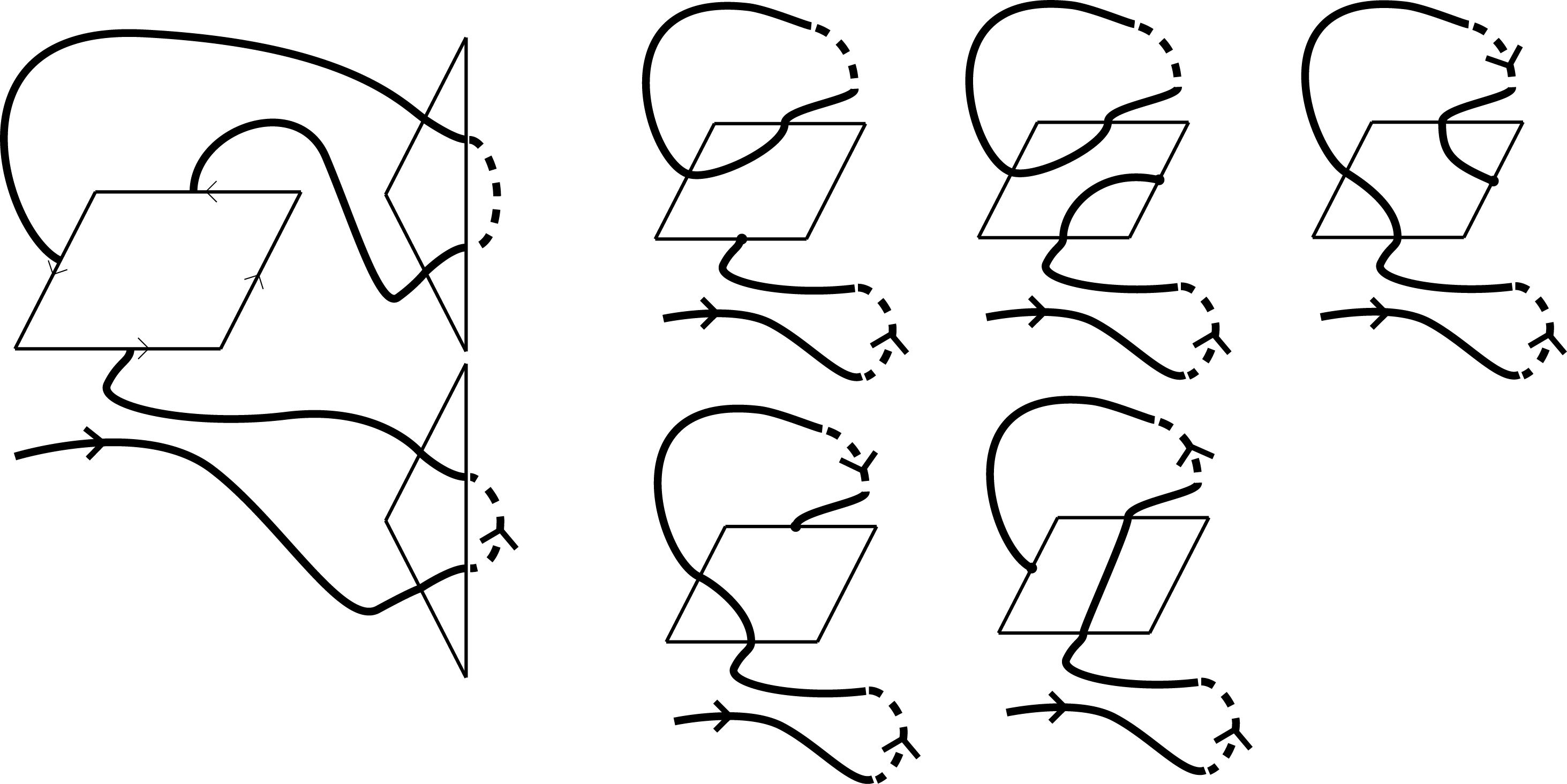}
\caption{Detailed configurations corresponding to the second connectivity in Fig.~\ref{fig:extcon3}.}
\label{fig:bulkblobblob}
\end{figure}

\begin{figure}[ht]
\centering
\includegraphics[scale=0.2]{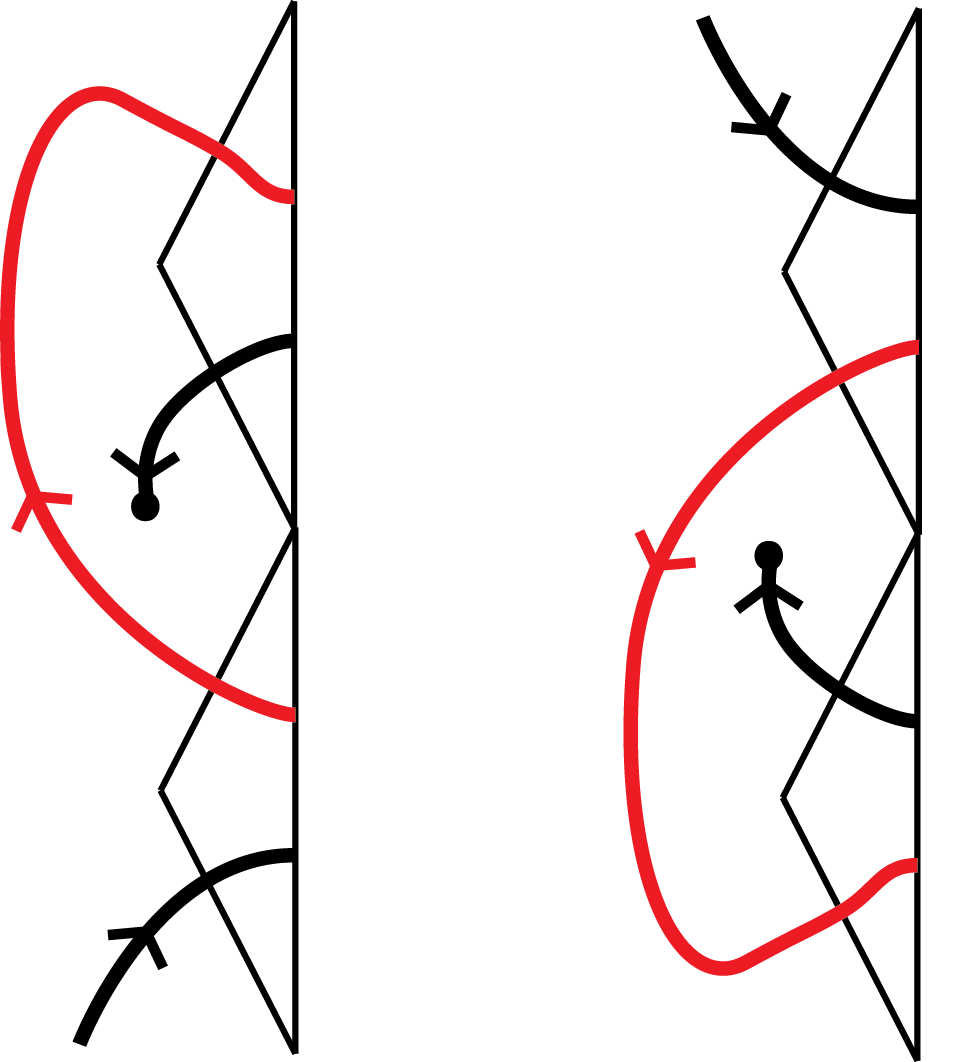}
\caption{The third type of boundary loop shown in red. It carries the weight $n_3$. }
\label{fig:n3}
\end{figure}

A second set of additional configurations occurs if the defect segment has passed through a $\beta_3$ plaquette \textit{before} entering the bulk plaquette, see Fig.~\ref{fig:extcon3}. To illustrate this, the configurations corresponding to the second connectivity in Fig.~\ref{fig:extcon3} are explicitly depicted in Fig.~\ref{fig:bulkblobblob}.
In this case, loops with weight $n_2$ may occur as well as a new type of boundary loop: a loop that passes from the top edge of a $\beta_3$ plaquette to the top edge of another $\beta_3$ plaquette, as shown on the left in Fig.~\ref{fig:n3}. This type of loop is \textit{only} possible because of the presence of the defect loop segment. We give such a loop the weight $n_3$. Similarly, a boundary loop from the bottom edge of a $\beta_3$ plaquette to the bottom edge of another $\beta_3$ plaquette is also given the weight $n_3$. 
There are yet more sets of configurations that can occur. For example, the loop segment may have passed through a $\beta_3$ plaquette before entering the bulk plaquette. There are also configurations where the external loop configuration passes multiple times through a $\beta_3$ plaquette. In all of these cases, however, one obtains equations that are equal, up to overall factors, to those derived from the configurations already considered above.

For these reasons, we modify the definition of the observable to
\be
 F(z) := \sum_{\gamma:\,-\infty \to z} P_{\mathrm{bdy}}(\gamma)\e^{-\ii s W(\gamma)} 
  (q_1)^{\epsilon}(\overline{q}_1)^{\bar{\epsilon}},\qquad 
   \epsilon,\bar{\epsilon},\epsilon+\bar{\epsilon}\in\{0,1\},
\label{eq:Fdefb}
\ee
where the Boltzmann weight $P_{\mathrm{bdy}}(\gamma)$ is defined similarly to the expression for the weight $P(\gamma)$ of the configurations in (\ref{eq:Fdef}) (see (\ref{PG})), but may also include contributions from the $n_3$-type loops. The possible factor $q_1$ or $\overline{q}_1$ is associated to the defect if it passes through a $\beta_3$ boundary plaquette in a counterclockwise ($\epsilon=1$, $\bar{\epsilon}=0$) or clockwise ($\epsilon=0$, $\bar{\epsilon}=1$) manner, respectively. We note that $\bar{\epsilon}$ and $\epsilon$ are separate variables, not complex conjugates of each other as the notation might otherwise suggest. If the defect passes through more than one $\beta_3$ plaquette, it is the orientation (counterclockwise or clockwise) of its \textit{last} visit which determines the choice of factor $q_1$ or $\overline{q}_1$.

For the cases in Fig.~\ref{fig:extcon2}, we thus obtain the following equations
\be
\begin{array}{rcl}
n_1 u_1+\overline{q}_1\zeta\xi^2 v - q_1 \overline{\xi} u_2-\zeta (q_1 w_2 + n_1 w_1) &=& 0, \\[.1cm]
n_1 u_2 + \zeta \xi (n_1 w_2 + \overline{q}_1 w_1) - \overline{q}_1 \xi u_1 - q_1\zeta \overline{\xi} v &=& 0, \\[.1cm]
n_1 v+ \overline{q}_1\zeta \xi^2 u_1-(\overline{q}_1\xi^2 w_1+q_1 \overline{\xi}^2 w_2) - q_1\zeta \overline{\xi} u_2 &=& 0.
\end{array}
\ee
Writing
\be
q_1=\rho\, \e^{4\ii\lambda+2\ii\lambda_1},\qquad n_1=-2\rho \cos2\lambda_1,
\ee
it is easy to verify that these are satisfied for arbitrary $\rho$ with the integrable Boltzmann weights given in \eqref{eq:BoltzWeightsBulk}. The factor $\rho$ is fixed below.
 
For the cases in Fig.~\ref{fig:extcon3}, we obtain the equations
\be
\begin{array}{rcl}
q_1n_1u_1+\zeta\xi^2\overline{q}_1n_3 v-\overline{\xi} q_1n_2u_2-\zeta q_1(n_2w_2 + n_1 w_1) &=& 0, \\[.1cm]
q_1n_1u_2+\zeta\xi (q_1n_1w_2+\overline{q}_1n_3w_1)-\xi \overline{q}_1 n_3 u_1 -\zeta\overline{\xi}q_1n_2v &=& 0, \\[.1cm]
q_1n_1 v+\zeta\xi^2\overline{q}_1 n_3 u_1 -(\xi^2\overline{q}_1 n_3 w_1 +\overline{\xi}^2q_1n_2w_2)-\zeta\overline{\xi} q_1n_2u_2 &=& 0.
\end{array}
\ee
Using the expressions \eqref{eq:BoltzWeightsBulk} for the Boltzmann weights, we then obtain a condition for $n_3$:
\be
 \e^{8\lambda\ii}\,\overline{q}_1(n_3-q_1)+q_1(n_2-q_1) = 0.
\ee
Since $n_3$ is the weight of a loop, we impose the additional constraint that it is real. This results in 
\be
 n_3=-n_2\frac{\sin 4\lambda}{\sin (4\lambda+4\lambda_1)},\qquad \rho=n_2 \frac{\sin 2 \lambda_1}{\sin (4\lambda+4\lambda_1) },
\ee
or equivalently,
\be
 \frac{n_1}{n_2} = -\frac{\sin 4\lambda_1}{\sin (4\lambda+4\lambda_1)},\qquad \frac{n_1}{n_3} = \frac{\sin 4\lambda_1}{\sin 4\lambda},
\ee
from which it follows that the various loop weights are related by 
\be
 n_3^2 = n_1^2+n_2^2-nn_1n_2.
\label{eqn:nnn}
\ee
We will comment on this relation in the discussion following (\ref{betannn}).

\section{Discrete holomorphicity at the boundary}
\label{Sec4}

Recall that, in the complex plane, we can write an integral of an analytic function $F=F_1+\ii F_2$ over a contour $C$ as the sum of two real-valued line integrals in $\mathbb{R}^2$,
\begin{equation}
\int_C F\dd z = \int_C \vec{F}\cdot \dd\vec{l} + \ii \int_C \vec{F}\cdot \dd\vec{n},
\label{eqn:line}
\end{equation} 
where $\vec{F}=(F_1,-F_2)$, $\dd \vec{l}$ is the infinitesimal line element parallel to $C$, while $\dd \vec{n}$ is the infinitesimal outward normal to $C$. If $C$ is a closed contour, holomorphicity requires both the real and imaginary parts in \eqref{eqn:line} to vanish, which we can think of as a zero-flux condition. However, if we integrate along the boundary of a domain, $C$ is not a closed contour. In this case, we will instead require \textit{either} the real part \textit{or} the imaginary part of \eqref{eqn:line} to vanish, corresponding to requiring that the perpendicular or parallel component of the flux vanishes.

In the discrete setting, we therefore have two natural choices of boundary conditions for the parafermionic observable:
\be
 \mathrm{Re}\, \Big( \sum_{i}F(z_i)\Delta z_i \Big) =0,
\label{eqn:bc1}
\ee
if we require zero flux of $\vec{F}$ \textit{along} the boundary, or
\be
 \mathrm{Im}\, \Big( \sum_{i}F(z_i)\Delta z_i \Big) =0,
\label{eqn:bc2}
\ee
if we require zero flux of $\vec{F}$ \textit{across} the boundary. The sum in \eqref{eqn:bc1} and \eqref{eqn:bc2} is along the two edges of a single boundary plaquette, and counterclockwise orientation means along the edges in the downward direction. We will now show that requiring these boundary conditions separately (not simultaneously) results in integrable boundary weights, i.e. weights satisfying the (boundary Yang-Baxter) reflection equation~\cite{Sklyanin}.
Throughout, we shall assume that the boundary plaquette weights $\beta_1$, $\beta_2$ and $\beta_3$ are real.

\subsection{Boundary configurations}

The first set of boundary configurations to consider are the ones where the defect loop segment enters a boundary plaquette without having passed through a $\beta_3$ plaquette prior to arriving at the boundary plaquette under consideration. 
\begin{figure}[ht]
\centering
\includegraphics[scale=0.25]{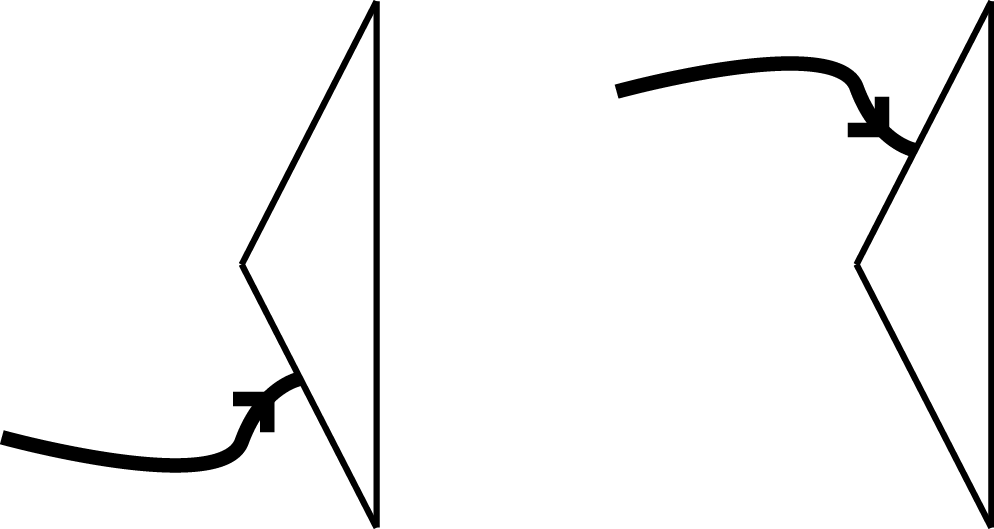}
\caption{First set of boundary configurations.}
\label{fig:boundary1}
\end{figure}
When entering from below, as indicated to the left in Fig.~\ref{fig:boundary1}, the expression for the discrete contour integral is
\be
 \sum_{i}F(z_i)\Delta z_i=\xi^{-1/2} \left ( \overline{\zeta} \beta_1-\zeta \beta_2-\zeta q_1\beta_3 \right).
\label{eqn:bcset1}
\ee
As we shall require that the observable satisfies the boundary condition \eqref{eqn:bc1} or \eqref{eqn:bc2}, let us compute the real ($R$) and imaginary ($I$) parts of the above expression:
\bea
 R_1 &=& \sin(x-\tfrac32\lambda)\beta_1+\sin(x+\tfrac32\lambda)\beta_2+\rho\sin(x-\tfrac52\lambda-2\lambda_1)\beta_3,\\[.1cm]
 I_1 &=& -\cos(x-\tfrac32\lambda)\beta_1+\cos(x+\tfrac32\lambda)\beta_2+\rho\cos(x-\tfrac52\lambda-2\lambda_1)\beta_3.
\label{eqn:bceqn1}
\eea

If the defect enters from above instead, we simply get minus the complex conjugate of \eqref{eqn:bcset1}. Since our boundary condition, (\ref{eqn:bc1}) or (\ref{eqn:bc2}), amounts to setting either the real or the imaginary part to zero, the choice of entry edge is immaterial. This observation also applies to (\ref{eqn:bcset2}) and (\ref{eqn:bcset3}) below.

As indicated in Fig.~\ref{fig:boundary2}, the second type of configurations corresponds to the defect entering the boundary plaquette after having already passed through a $\beta_3$ boundary plaquette. For the case to the left in Fig.~\ref{fig:boundary2}, we find
\be
 \sum_{i}F(z_i)\Delta z_i
  =\xi^{-1/2}q_1\left(\overline{\zeta}\beta_1-\zeta \beta_2-\zeta n_2\beta_3 \right),
\label{eqn:bcset2}
\ee
whose real and imaginary parts are given by 
\bea
R_2&=&\rho\,\big[\sin(x+\tfrac52\lambda+2\lambda_1)\beta_1+\sin(x-\tfrac52\lambda-2\lambda_1)(\beta_2+n_2\beta_3)\big],\\[.1cm]
I_2&=&\rho\,\big[-\cos(x+\tfrac52\lambda+2\lambda_1)\beta_1+\cos(x-\tfrac52\lambda-2\lambda_1)(\beta_2+n_2\beta_3)\big].
\label{eqn:bceqn2}
\eea

\begin{figure}[ht]
\centering
\includegraphics[scale=0.25]{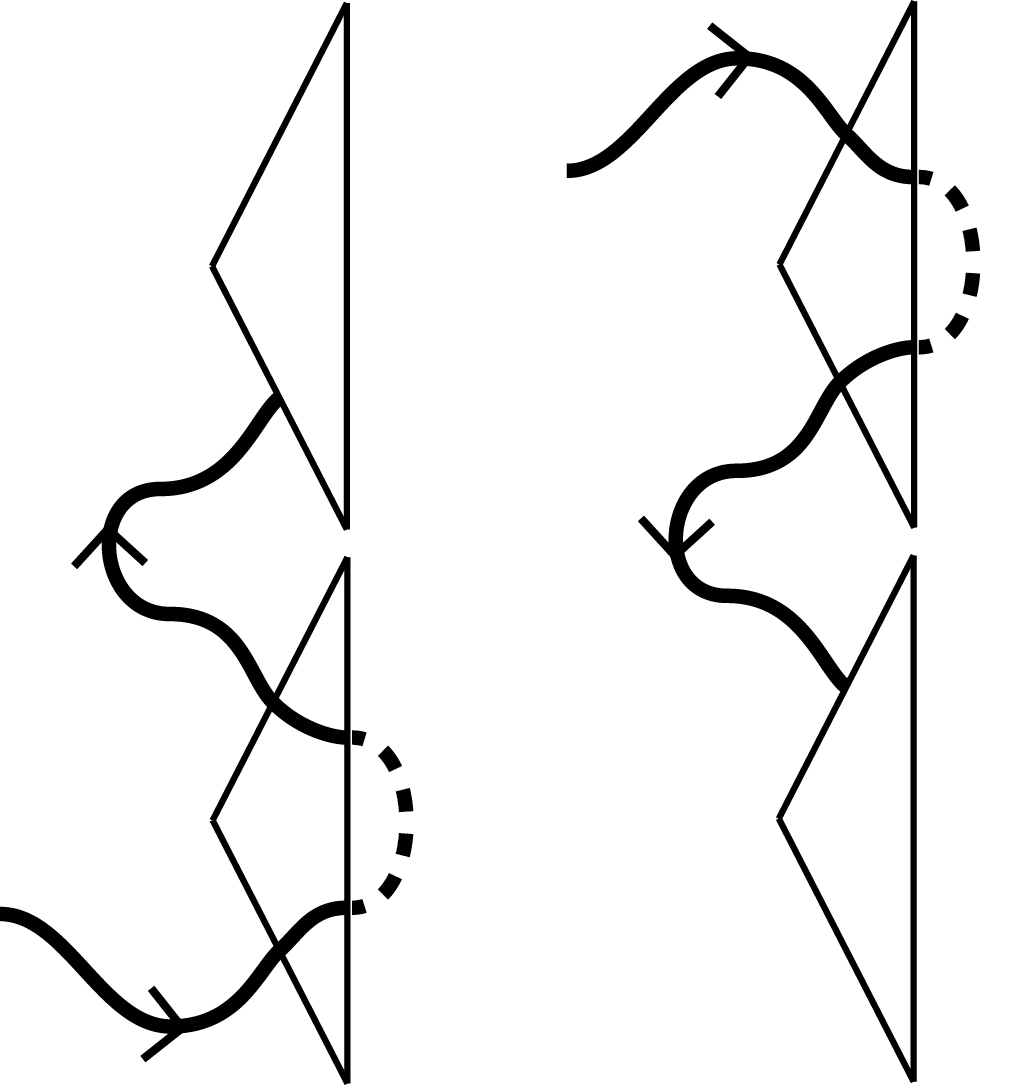}
\caption{Second set of boundary configurations.}
\label{fig:boundary2}
\end{figure}
\begin{figure}
\centering
\includegraphics[scale=0.25]{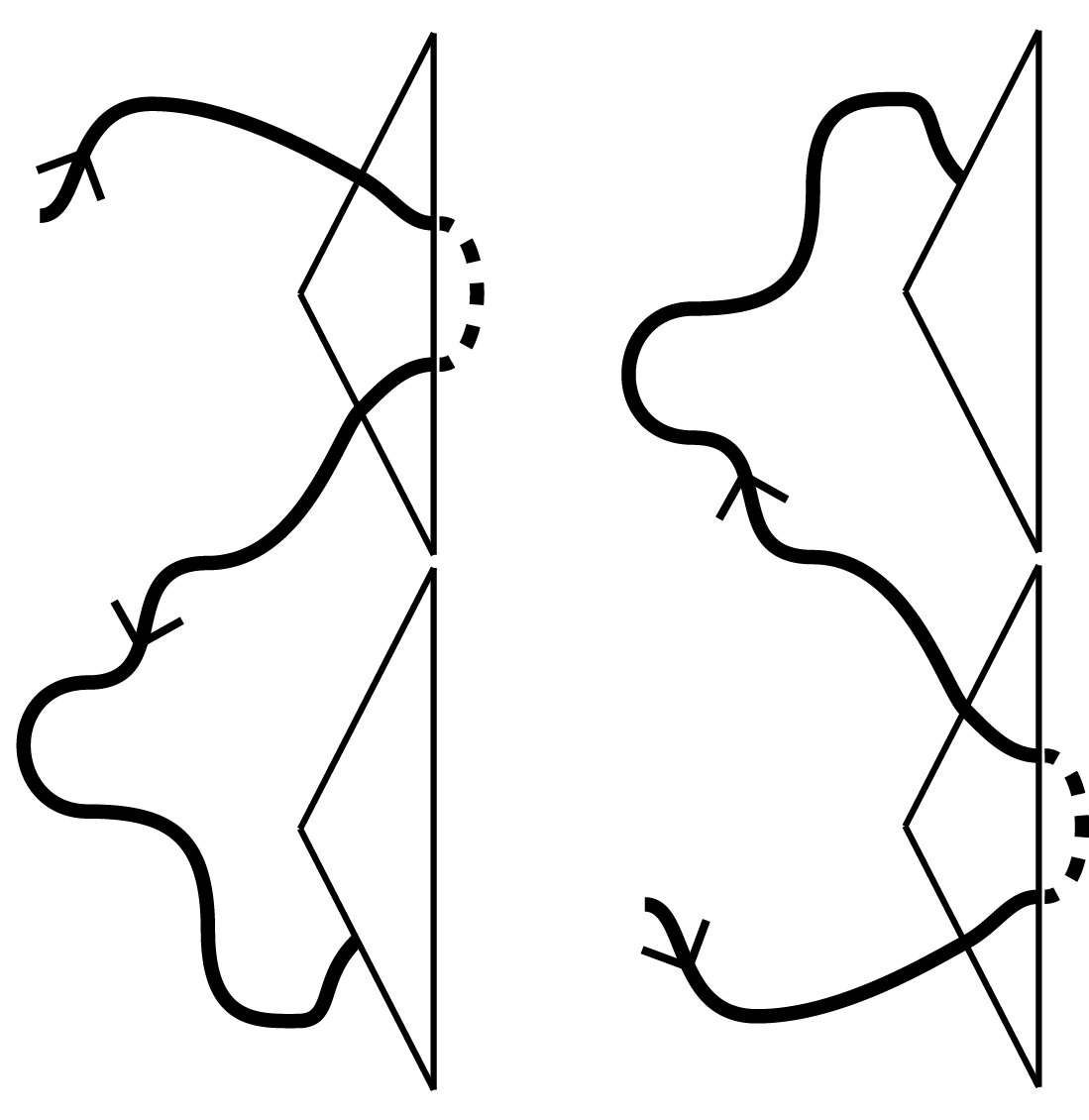}
\caption{Third set of boundary configurations.}
\label{fig:boundary3}
\end{figure}

The final set of boundary configurations to consider involves those where the defect loop passes clockwise through a $\beta_3$ boundary plaquette before entering the boundary plaquette under consideration in a counterclockwise manner (from below), or vice-versa. As follows from Fig.~\ref{fig:boundary3}, each of these cases includes the possibility of an $n_3$-type loop. 
Indeed, the discrete contour integral for the case to the left in Fig.~\ref{fig:boundary3} is given by
\be
 \sum_{i}F(z_i)\Delta z_i
  =\xi^{-1/2} \left (\overline{\zeta}\,\overline{q}_1 \beta_1-\zeta\,\overline{q}_1 \beta_2-\zeta q_1 n_3 \beta_3 \right),
\label{eqn:bcset3}
\ee
whose real and imaginary parts are given by
\bea
 R_3&=&\rho\,\big[\sin(x-\tfrac{11}2\lambda-2\lambda_1)\beta_1+\sin(x+\tfrac{11}2\lambda+2\lambda_1)\beta_2+n_3 \sin(x-\tfrac52\lambda-2\lambda_1)\beta_3\big],\\[.1cm]
 I_3&=&\rho\,\big[-\cos(x-\tfrac{11}2\lambda-2\lambda_1)\beta_1+\cos(x+\tfrac{11}2\lambda+2\lambda_1)\beta_2+n_3 \cos(x-\tfrac52\lambda-2\lambda_1)\beta_3\big].
\label{eqn:bceqn3}
\eea

Setting $R_1=R_2=R_3=0$ yields three equations, one of which is linearly dependent on the other two, allowing us to solve for the ratios $\beta_1(x) / \beta_3(x)$ and $\beta_2(x) / \beta_3(x)$. Likewise, we can solve for the same ratios in the case $I_1=I_2=I_3=0$. This gives the solution sets
\be
\begin{array}{rcl}
 \beta_1(x) &=& n_3 \left[ \pm\cos(2x-\lambda) - \cos (4\lambda+4\lambda_1) \right],\\[.1cm]
 \beta_2(x) &=& n_3\left[\pm\cos\lambda - \cos(2x - 4\lambda - 4\lambda_1) \right], \\[.1cm]
 \beta_3(x) &=& -2\sin 4\lambda\,\sin 2x,
\end{array}
\ee
where the $+$ sign is for vanishing \textit{real} parts, while the $-$ sign is for vanishing \textit{imaginary} parts. Written in terms of the boundary loop fugacities $n_1$, $n_2$ and $n_3$ explicitly, the weights read
\be
\begin{array}{rcl}
 \beta_1(x) &=& n_1 + n_2\cos 4\lambda \pm n_3 \cos(2x-\lambda), \\[.1cm]
 \beta_2(x) &=& n_1\cos 2x + n_2\cos(2x-4\lambda) \pm n_3 \cos \lambda, \\[.1cm]
 \beta_3(x) &=& -2\sin 4\lambda\,\sin 2x,
\end{array}
\label{betannn}
\ee
where we recall the relation $n_3^2 = n_1^2+n_2^2-nn_1n_2$ given in (\ref{eqn:nnn}). 

We have verified explicitly that the weights (\ref{betannn}) precisely match the integrable boundary plaquette weights following from solving the (boundary Yang-Baxter) reflection equation for this model. From that perspective, $n_3$ is merely a parameter defined by (\ref{eqn:nnn}) since $n_3$-type loops can \textit{not} be formed without the defect.
One can generalise the model, though, by introducing boundary plaquettes with a \textit{single} loop segment, thereby allowing $n_3$-type loops to be formed. These new boundary plaquettes and the solution to the corresponding reflection equation are discussed in Appendix~\ref{AppAsymmOn}. Whether a natural parafermionic observable, compatible with discrete holomorphicity, can be defined in this generalised $O(n)$ model remains to be seen.

In the special case
\be
 n_1=-\frac{\sin4\lambda_1}{\sin(4\lambda+4\lambda_1)},\qquad n_2=1,
\label{spec}
\ee 
the boundary plaquette weights (\ref{betannn}) reduce to 
\be
\begin{array}{rcl}
 \hat{\beta}_1(x) &=&\tfrac{1}{2}[\cos(2x-\lambda)\mp\cos(4\lambda+4\lambda_1)], \\[.1cm]
 \hat{\beta}_2(x) &=&\tfrac{1}{2}[\cos\lambda\mp\cos(2x-4\lambda-4\lambda_1)], \\[.1cm]
 \hat{\beta}_3(x) &=& \pm\sin(4\lambda+4\lambda_1)\,\sin 2x,
\end{array}
\label{betahat}
\ee
where we have introduced the sign-dependent rescaling
\be
 \hat{\beta_j}(x)=\mp\frac{\sin(4\lambda+4\lambda_1)}{2\sin4\lambda}\beta_j(x),\qquad j=1,2,3.
\ee
In factorized form, these solutions are recognised as the ones appearing in the $\emph{blobbed}$ $O(n)$ model of Dubail, Jacobsen and Saleur~\cite{DJSb}. 
Even though the relation $n_3^2 = n_1^2+n_2^2-nn_1n_2$ is homogeneous in the boundary loop fugacities ($n_1$, $n_2$ and $n_3$), the specialisation (\ref{spec}) is not merely a rescaling (assuming $n_2\neq0$) since $\beta_3$ is not proportional to any of the loop fugacities. Our solutions are therefore more general than the ones in~\cite{DJSb}.

The diagonal or reflecting boundary conditions considered by Ikhlef~\cite{Ikhlef} correspond to setting $\beta_3=0$, in which case we should only consider the boundary configurations depicted in Fig.~\ref{fig:boundary1}. Imposing the boundary condition $R_1=0$ yields
\be
 \beta_1(x)=\sin(\tfrac{3}{2}\lambda+x),\qquad \beta_2(x)=\sin(\tfrac{3}{2}\lambda-x),
 \label{eqn:ondbc}
\ee
while imposing the boundary condition $I_1=0$ yields
\be
 \beta_1(x)=\cos(\tfrac{3}{2}\lambda+x),\qquad\beta_2(x)=\cos(\tfrac{3}{2}\lambda-x).
\ee
As solutions to the reflection equation, these two solutions were originally obtained by Batchelor and Yung~\cite{Batchelor}.

\section{The \boldmath{$C_2^{(1)}$} loop model}
\label{Sec5}

The $C_2^{(1)}$ loop model is a model of densely packed loops of two possible colours. In the bulk, a parafermionic observable has been defined by Ikhlef and Cardy~\cite{IkhlefCardy}. After reviewing this definition in Section~\ref{SecDefTwo}, we extend it to accommodate non-trivial boundary conditions. 

\begin{figure}[ht]
\centering
\begin{picture}(150,225)
\put(0,0){\includegraphics[scale=0.2]{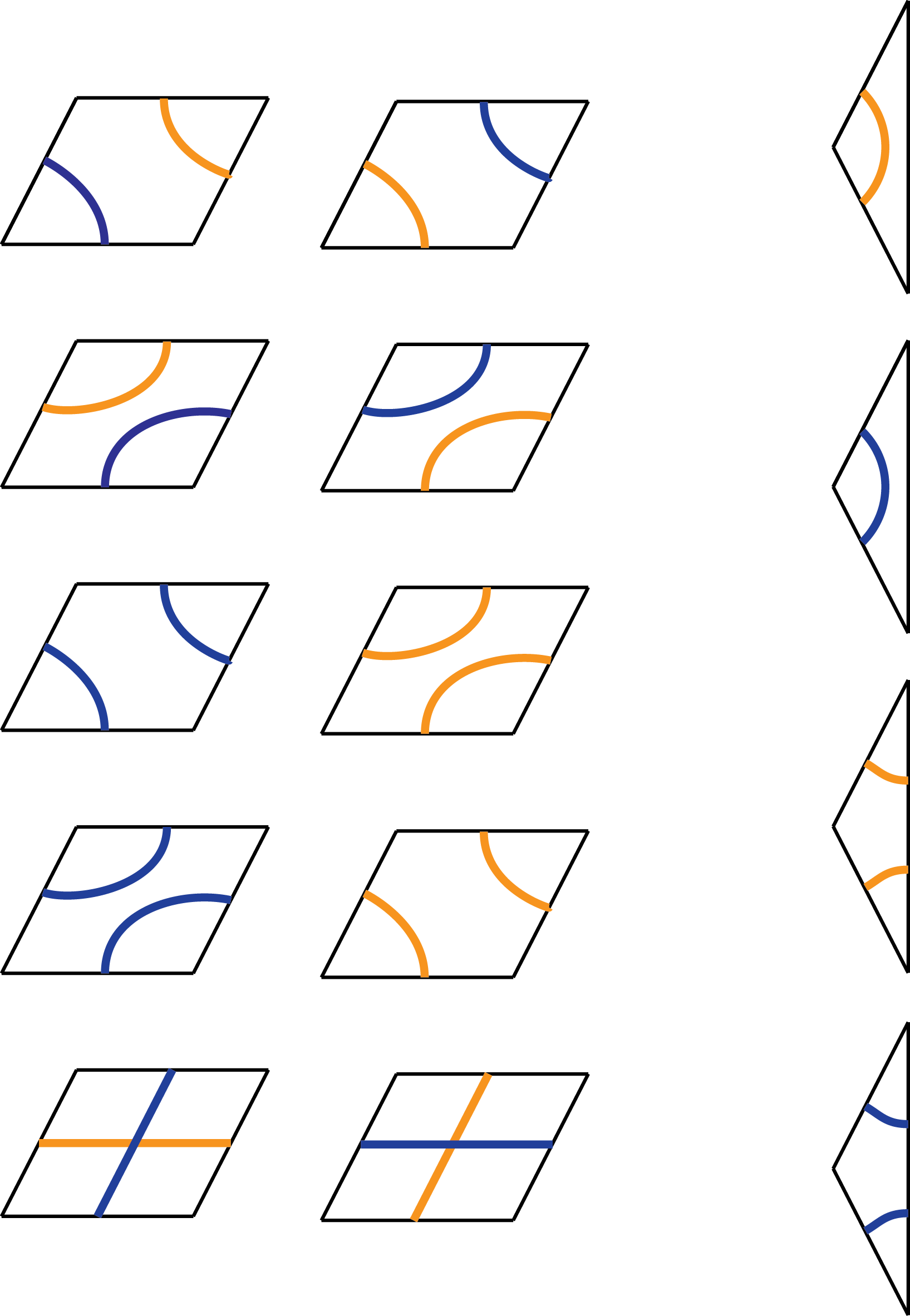}}
\put(130, 200){$\beta_1$}
\put(130, 142){$\beta_2$}
\put(130, 84){$\beta_3$}
\put(130, 26){$\beta_4$}
\put(-15, 200){$u_1$}
\put(-15, 158){$u_2$}
\put(-15, 115){$w_1$}
\put(-15, 75){$w_2$}
\put(-15, 30){$v$}
\end{picture}
\caption{Bulk and boundary plaquettes of the $C_2^{(1)}$ loop model. The inclusion of the $\beta_3$ and $\beta_4$ plaquettes gives rise to loop types not present in the bulk version of the model.}
\label{fig:c21boundary}
\end{figure}

The set of bulk and boundary plaquettes are shown in Fig.~\ref{fig:c21boundary}. There are also three types of loops in the model. Loops that lie entirely within the bulk have fugacity $n$. As in the dilute $O(n)$ model, loops that are attached to the boundary are of two types: those whose top end is attached to the $top$ edge of a boundary plaquette have fugacity $n_1$, while those whose top end is attached to the $lower$ edge of a boundary plaquette have fugacity $n_2$. The loop fugacities $n$ and $n_1$ are parameterised as
\be
 n=-2\cos 4\lambda, \qquad n_1=-2\rho\cos 2\lambda_1.
\label{nn1}
\ee
The parameterisation of $n_2$, as well as how $\rho$ is determined, is discussed below.
%

\subsection{Parafermionic observable}
\label{SecDefTwo}

\begin{figure}
\centering
\includegraphics[scale=0.2]{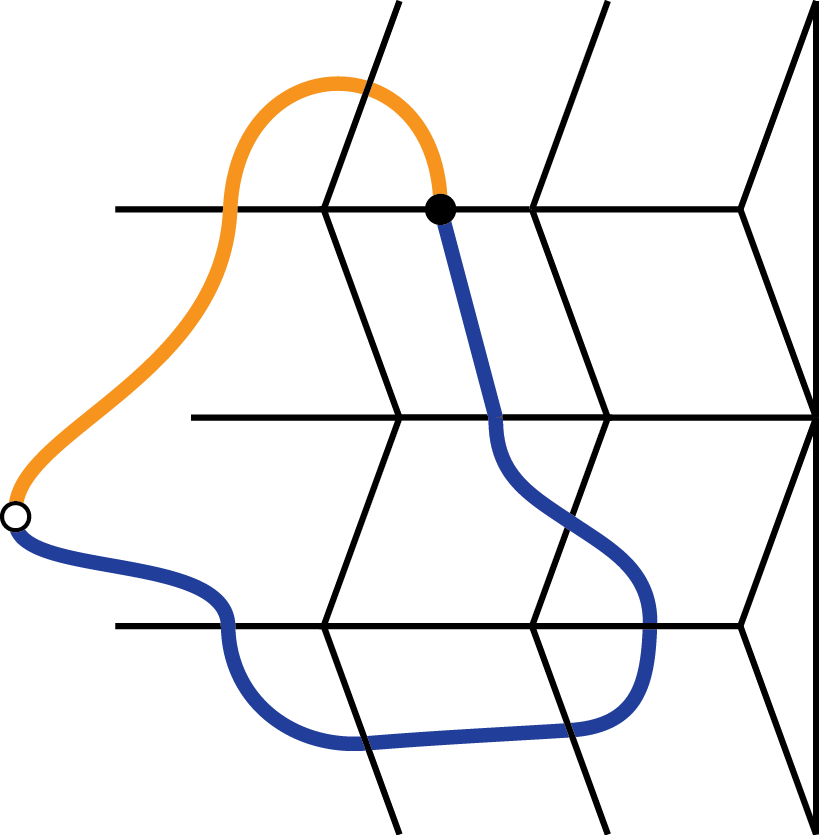}
\caption{The definition by Ikhlef and Cardy~\cite{IkhlefCardy} of the parafermionic observable $F(z)$ in the $C_2^{(1)}$ loop model. The open circle represents a point at $-\infty$, while the black disk is the point $z$ at which the two colours meet.}
\label{fig:c21parafermion}
\end{figure}
We again introduce a defect in the model, this time consisting of a closed loop formed by gluing together two open loop segments of different colours, see Fig.~\ref{fig:c21parafermion}. The value of the parafermionic observable $F(z)$ is defined at the point $z$ where the colour changes,
\be
F(z) := \sum_{\gamma:-\infty \rightarrow z}P(\gamma)\e^{-\ii s W(\gamma)}.
\label{eqn:Fdefc21}
\ee
The Boltzmann weight $P(\gamma)$ of the configuration $\gamma$ is defined as in the $O(n)$ model as the product of the plaquette and loop weights.
The winding angle $W(\gamma)$ is the \textit{sum} of the winding angles of the two loop segments of the defect, where the initial orientations of the segments are along the positive and negative imaginary axes, respectively.

As before, the above definition has to be modified in the boundary model because of possible boundary interactions of the defect loop. A configuration acquires a factor $q_1$ ($\overline{q}_1$) if the segment passes through the boundary in the counterclockwise (clockwise) sense. In certain configurations, the loop segment passes through the boundary more than once. Such configurations acquire a factor according to how the segment passes through the boundary the \textit{final} time. 

Owing to the presence of the defect loop, boundary loops can be formed between two upper edges or between two lower edges of a pair of boundary plaquettes. We assign such boundary loops the fugacity $n_3$. As in the dilute $O(n)$ model (\ref{eq:Fdefb}), we thus define the parafermionic observable by
\be
 F(z) := \sum_{\gamma:\,-\infty \to z} P_{\mathrm{bdy}}(\gamma)\e^{-\ii s W(\gamma)} 
  (q_1)^{\epsilon}(\overline{q}_1)^{\bar{\epsilon}},\qquad 
   \epsilon,\bar{\epsilon},\epsilon+\bar{\epsilon}\in\{0,1\},
\ee
where $P_{\mathrm{bdy}}(\gamma)$ is the weight of the configuration $\gamma$ which may involve $n_3$-type loops.

The parameters $q_1$, $n_2$ and $n_3$ are fixed from requiring discrete holomorphicity in the bulk and that $n_3$ be real. Here we summarise the results which are verified in Section~\ref{SecC21Bulk}. Thus, as in the dilute $O(n)$ model, $q_1$, $n_2$ and $n_3$ are conveniently parameterised as
\be
q_1 = -\frac{n_1}{2\cos2\lambda_1}\e^{\ii(4\lambda+2\lambda_1)}, \qquad n_2 = -n_1\frac{\sin(4\lambda+4\lambda_1)}{\sin4\lambda_1}, \qquad n_3=n_1\frac{\sin 4\lambda}{\sin4\lambda_1}.
\label{eqn:c21factors}
\ee

\subsection{Bulk configurations}
\label{SecC21Bulk}

The integrable bulk weights~\cite{WarnaarNienhuis} are

\be
\begin{array}{rcl}
 u_1(x) &=& \sin2\lambda\sin(x-6\lambda), \\[.1cm]
 u_2(x) &=& -\sin2\lambda\sin x, \\[.1cm]
 v(x) &=& -\sin x \sin(x-6\lambda), \\[.1cm]
 w_1(x) &=& \sin(x-2\lambda)\sin(x-6\lambda), \\[.1cm]
 w_2(x) &=& -\sin x \sin(x-4\lambda).
\end{array}
\label{eq:BoltzWeightsBulkC21}
\ee
In the following, we use 
\be
 \xi=\e^{2\pi s \ii},\qquad \zeta=\e^{- \ii x},\qquad x=\alpha(2s-1),
\ee 
where $s=\frac{3\lambda}{2}-\frac{1}{2}$. 

\begin{figure}
\centering
\includegraphics[scale=0.2]{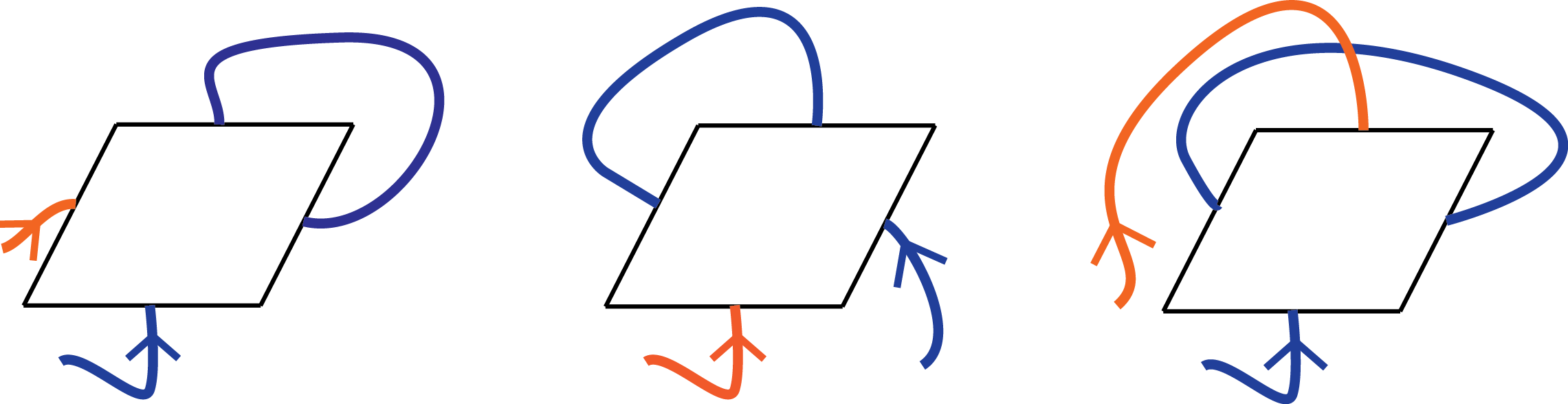}
\caption{The first set of bulk configurations in the $C_2^{(1)}$ model. These configurations  are the only ones appearing in the bulk model considered by Ikhlef and Cardy~\cite{IkhlefCardy}.}
\label{Fig:c21extcon1}
\end{figure}

\begin{figure}
\centering
\includegraphics[scale=0.2]{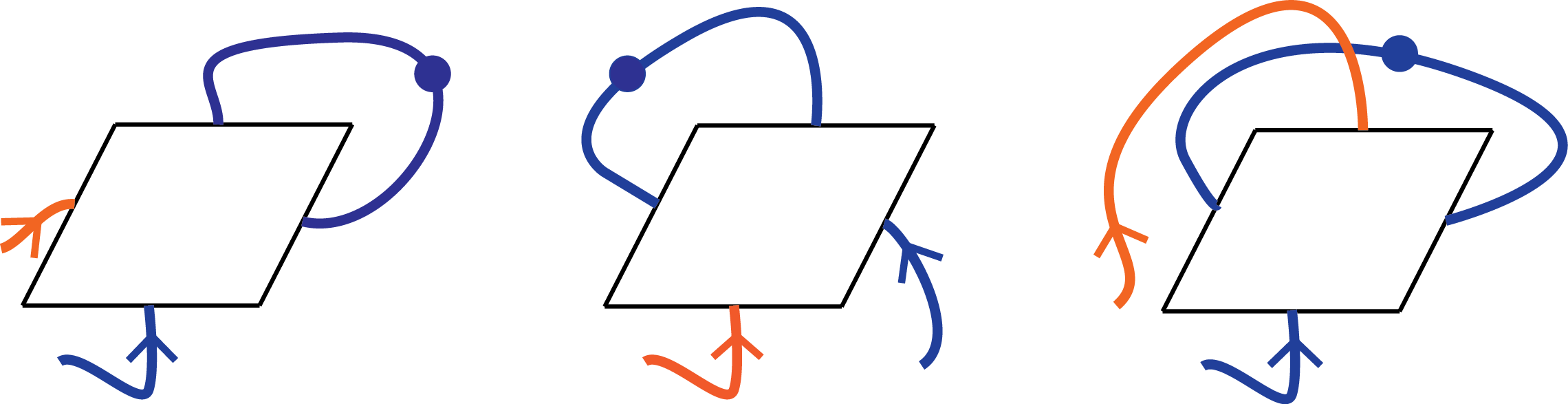}
\caption{The first additional set of bulk configurations in the $C_2^{(1)}$ model. In these cases, the defect passes through a $\beta_4$ plaquette, indicated with a blob, before re-entering the given bulk plaquette. Similar configurations with a $\beta_3$ plaquette follow by interchanging the two colours.}
\label{fig:c21extcon2}
\end{figure}

\begin{figure}
\centering
\includegraphics[scale=0.2]{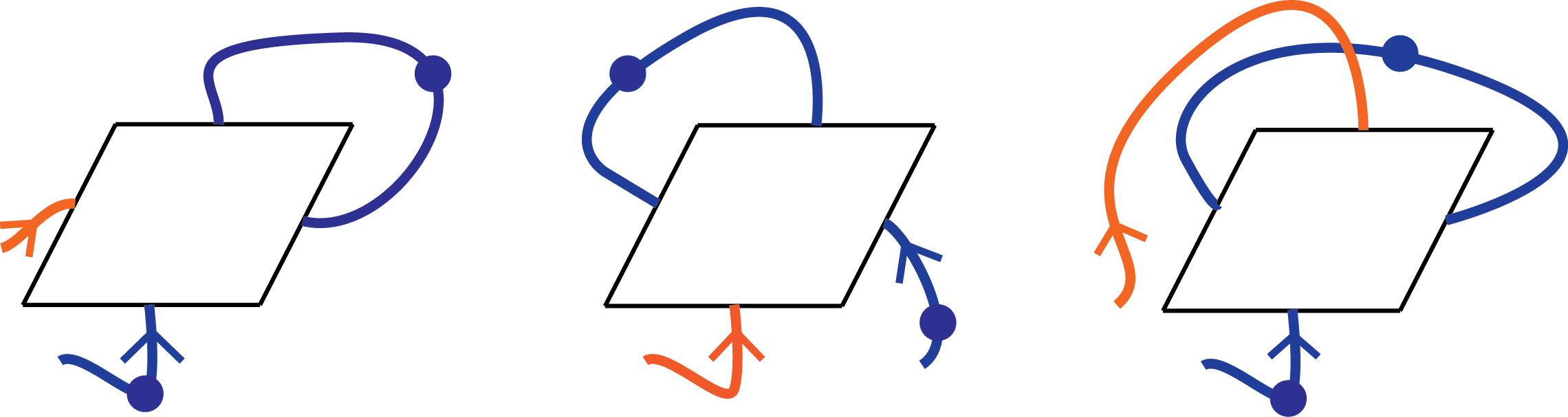}
\caption{The second additional set of bulk configurations in the $C_2^{(1)}$ model, here illustrated for a particular choice of colours. In these cases, the defect is blobbed before entering and possibly re-entering the bulk plaquette.}
\label{fig:c21extcon3}
\end{figure}

First, we consider the configurations which also appear in the bulk model. These are indicated in Fig.~\ref{Fig:c21extcon1}, and discrete holomorphicity gives the following equations
\be
\begin{array}{rcl}
 n u_1+\xi\zeta v- \overline{\xi}u_2-n\zeta w_1 - \zeta w_2 &=& 0, \\[.1cm]
 n\overline{\xi}w_2 + \overline{\xi}w_1 + n\zeta u_2 - v-\overline{\xi}\zeta u_1 &=& 0, \\[.1cm]
 nv + \xi\zeta u_1-\xi w_1 - \overline{\xi}w_2-\zeta u_2 &=& 0.
\end{array}
\ee
There are additional sets of bulk configurations, due to the possibility of the defect passing through the boundary via a $\beta_3$ or $\beta_4$ plaquette. In the first additional set, shown in Fig.~\ref{fig:c21extcon2}, the defect enters the given bulk plaquette before entering a $\beta_3$ or $\beta_4$ boundary plaquette. This gives the following three equations
\be
\begin{array}{rcl}
 n_1u_1+\overline{q}_1\xi\zeta v- q_1\overline{\xi}u_2-n_1\zeta w_1 - q_1\zeta w_2 &=& 0, \\[.1cm]
 n_1\overline{\xi}w_2 + q_1\overline{\xi}w_1 + n_1 \zeta u_2 - \overline{q}_1v-q_1\overline{\xi}\zeta u_1 &=& 0, \\[.1cm]
 n_1v + \overline{q}_1\xi\zeta u_1-\overline{q}_1\xi w_1 - q\overline{\xi}w_2-q_1\zeta u_2 &=& 0.
\end{array}
\ee
In the second set of additional configurations, shown in Fig.~\ref{fig:c21extcon3}, the defect passes through a $\beta_3$ or $\beta_4$ plaquette possibly twice: once before entering the given bulk plaquette and possibly again before re-entering. Requiring discrete holomorphicity imposes the conditions
\be
\begin{array}{rcl}
 n_1qu_1+n_3\overline{q}_1\xi\zeta v- n_2q_1\overline{\xi}u_2-n_1q\zeta w_1 - n_2q_1\zeta w_2 &=& 0, \\[.1cm]
n_1q_1\overline{\xi}w_2 + n_2q_1\overline{\xi}w_1 + n_1q_1\zeta u_2 - n_3\overline{q}_1v - n_2 q_1 \overline{\xi}\zeta &=& 0, \\[.1cm]
 n_1 q_1v + n_3\overline{q}_1\xi\zeta u_1-n_3\overline{q}_1\xi w_1 - n_2q\overline{\xi}w_2-n_2q\zeta u_2 &=& 0.
\end{array}
\ee
It is straightforward to verify that both sets of equations are satisfied using the parameterisations of $n_1$, $n_2$, $n_3$ and $q_1$ in (\ref{nn1}) and (\ref{eqn:c21factors}). This implicitly fixes the parameter $\rho$ in \eqref{nn1}.

\begin{figure}
\centering
\includegraphics[scale=0.2]{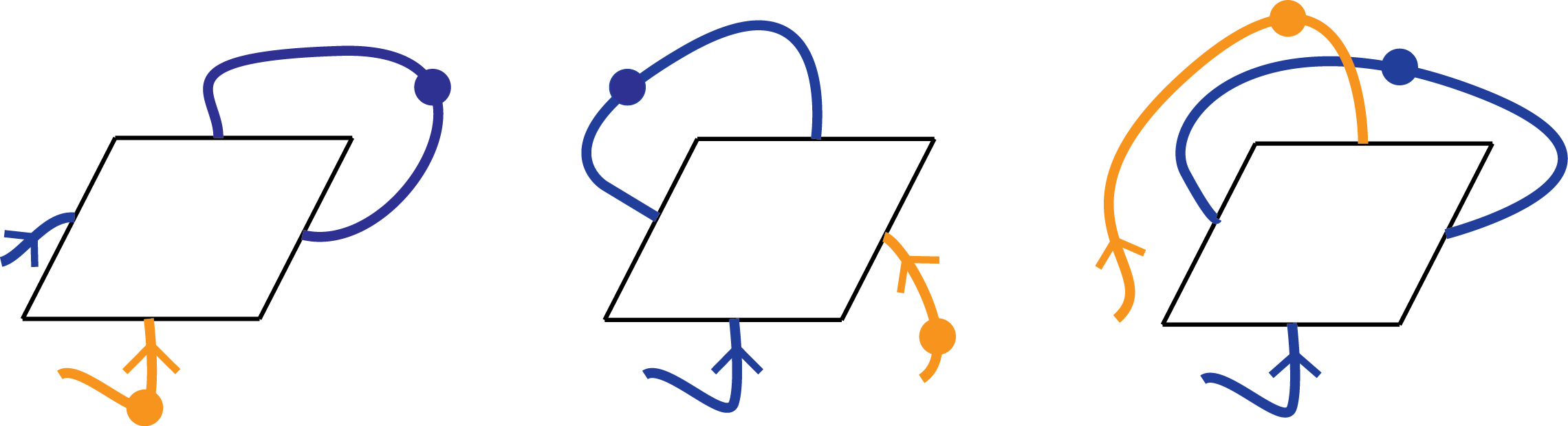}
\caption{Another set of additional bulk configurations in the $C_2^{(1)}$ model, here illustrated for a particular choice of colours. In these cases, both coloured segments of the defect are blobbed; one of them before entering the bulk plaquette.}
\label{fig:c21extcon4}
\end{figure}

There are yet more sets of configurations to consider. One set arises from interchanging the colours of those already considered. However, due to the bulk weights being symmetric under interchanging colours, the equations from these configurations are readily satisfied. Other sets arise from configurations where a loop segment, that cannot close to form a loop, is blobbed, see Fig.~\ref{fig:c21extcon4}. In all these cases, the ensuing equations are equivalent to the ones already considered.
%
\subsection{Boundary conditions and integrable weights}

\subsubsection{Diagonal boundaries}

\begin{figure}
\centering
\includegraphics[scale=0.3]{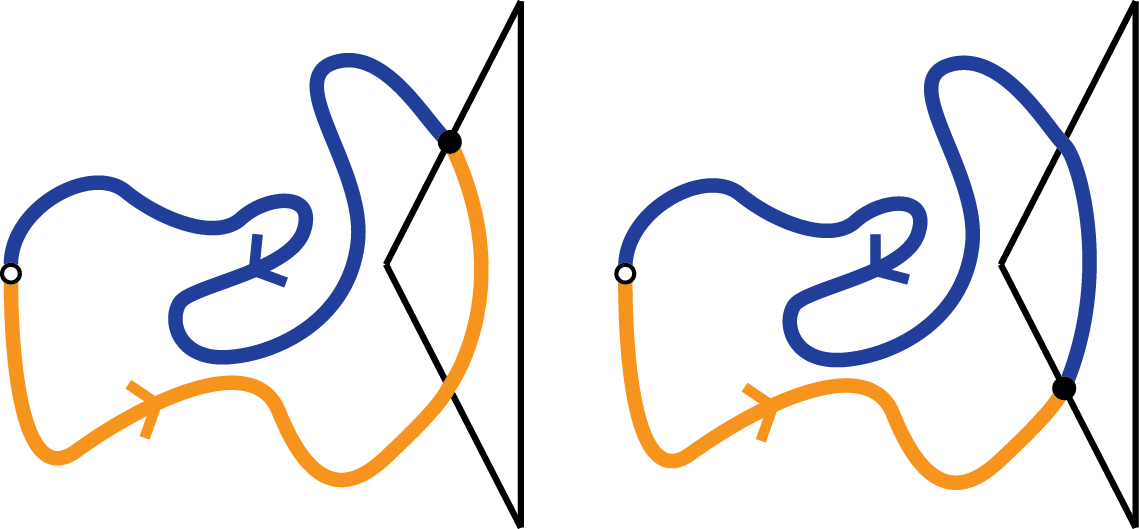}
\caption{Boundary configurations in the case of diagonal boundary plaquettes. The black disk indicates the point where the two defect loop segments are joined. The defect loop segments originate from $-\infty$, indicated here and in subsequent diagrams by an open circle. The blue (orange) loop segments are initially oriented along the positive (negative) imaginary axes. }
\label{fig:c21bc1}
\end{figure}

Diagonal boundary conditions correspond to $\beta_3 = \beta_4 = 0$, in which case we only need to consider the configurations in Fig.~\ref{fig:c21bc1}. The discrete contour integral is given by 
\be
  \sum_{i}F(z_i)\Delta z_i
  =\overline{\zeta}\beta_1-\zeta\beta_2.
\ee
The vanishing of the real or imaginary part gives, respectively,
\be
\beta_1 = \beta_2, \qquad \beta_1=-\beta_2.
\ee

\subsubsection{Non-diagonal boundaries}

\begin{figure}
\centering
\includegraphics[scale=0.3]{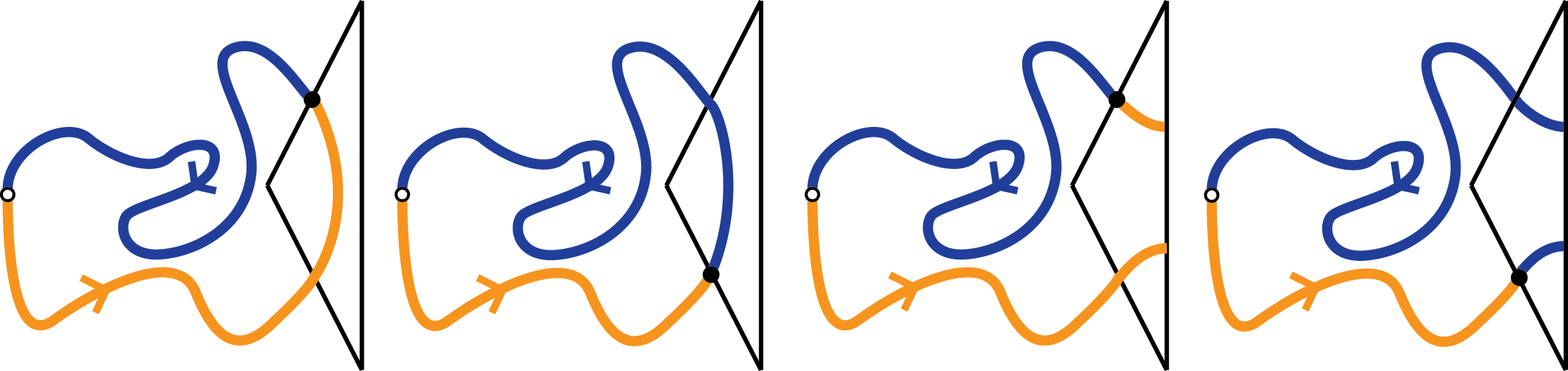}
\caption{The first set of configurations in the case of non-diagonal boundaries.}
\label{fig:c21bc2a}
\end{figure}

\begin{figure}
\centering
\includegraphics[scale=0.3]{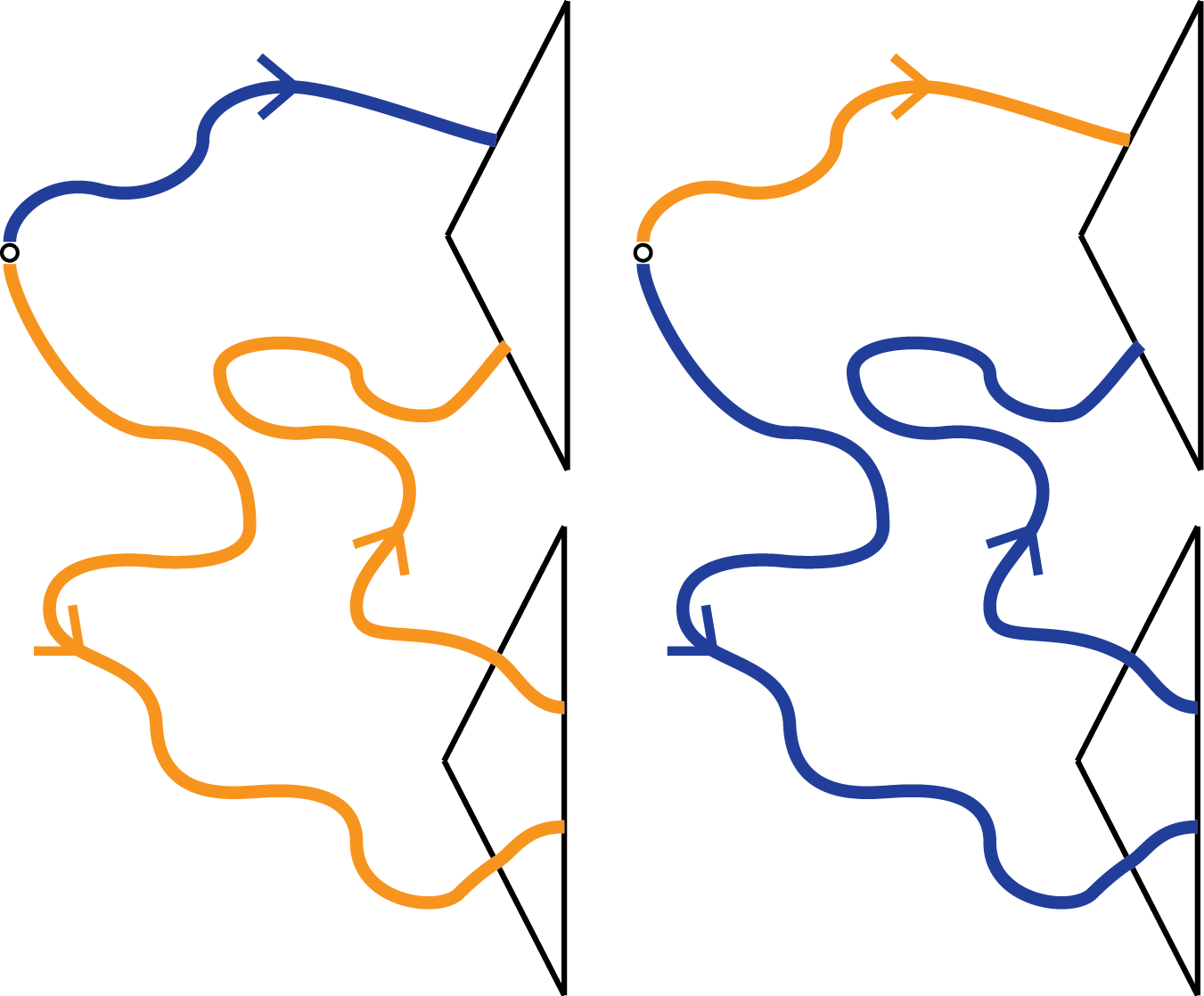}
\caption{The second and third set of configurations in the case of non-diagonal boundaries.}
\label{fig:c21bc2b}
\end{figure}

\begin{figure}
\centering
\includegraphics[scale=0.3]{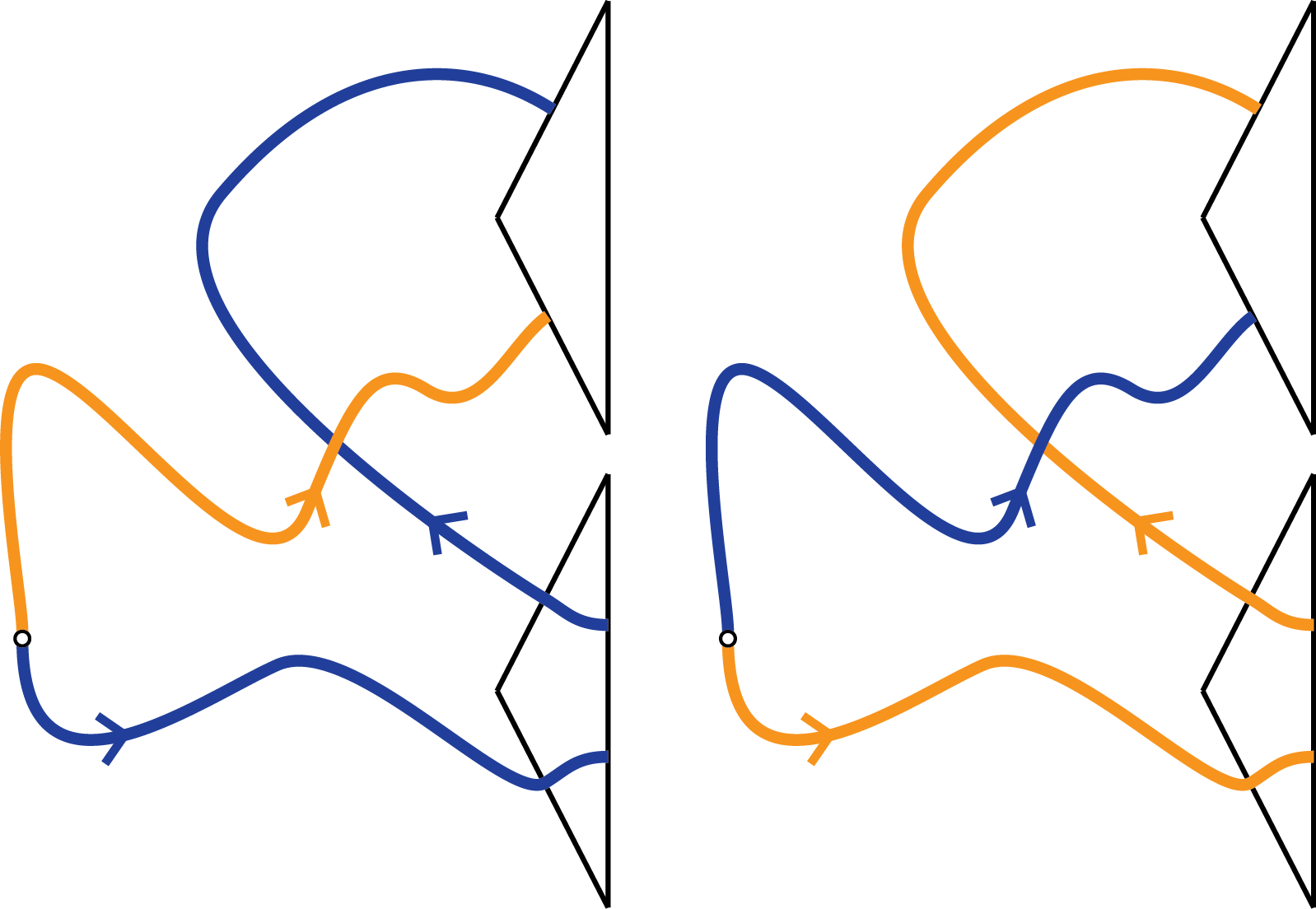}
\caption{The fourth and fifth set of configurations in the case of non-diagonal boundaries.}
\label{fig:c21bc2c}
\end{figure}

We now have four weights to solve for: $\beta_1, \beta_2, \beta_3$ and $\beta_4$, and five types of configurations to consider at the boundary. These are shown in Figs.~\ref{fig:c21bc2a}, \ref{fig:c21bc2b} and~\ref{fig:c21bc2c}. In the first set (Fig.~\ref{fig:c21bc2a}), the defect loop segments only pass through a single $\beta_3$ or $\beta_4$ plaquette. The discrete contour integral is
\be
  \sum_{i}F(z_i)\Delta z_i
  = -\zeta\beta_1+\overline{\zeta}\beta_2-q_1\zeta\beta_3+\overline{q}_1\overline{\zeta}\beta_4.
\label{eqn:c21bceqn1}
\ee
The real and imaginary parts of the above expression are
\bea
 R_1 &=& -\cos x\beta_1 + \cos x \beta_2 - \rho\cos(x-4\lambda-2\lambda_1)\beta_3 + \rho\cos(x-4\lambda-2\lambda_1)\beta_4, \\[.1cm]
 I_1 &=& \sin x \beta_1 + \sin x\beta_2 + \rho\sin(x-4\lambda-2\lambda_1)\beta_3 + \rho\sin(x-4\lambda-2\lambda_1)\beta_4,
\eea
where
\be
 \rho = -\frac{n_1}{2\cos2\lambda_1}.
\ee
In the second set, shown on the left in Fig.~\ref{fig:c21bc2b}, the orange loop segment has already passed through a $\beta_3$ plaquette. The two loop segments form a closed loop with total winding angle $2\pi$. The discrete contour integral yields
\be
  \sum_{i}F(z_i)\Delta z_i
  = -q_1\zeta\beta_1+q_1\overline{\zeta}\beta_2-q_1n_2\zeta\beta_3+\overline{q}_1 q_1 \overline{\zeta}\beta_4,
\label{eqn:c21bceqn2}
\ee
and its real and imaginary parts are
\bea
\!\!\!\!R_2 &\!\!\!\!=\!\!\!\!& -\rho\cos(x-4\lambda-2\lambda_1)\beta_1+\rho\cos(x+4\lambda+2\lambda_1)\beta_2-n_2\rho\cos (x-4\lambda-2\lambda_1)\beta_3+\rho^2\!\cos x\beta_4, \\[.1cm]
\!\!\!\!I_2 &\!\!\!\!=\!\!\!\!& \rho\sin(x-4\lambda-2\lambda_1)\beta_1+\rho\sin(x+4\lambda+2\lambda_1)\beta_2+n_2\rho\sin (x-4\lambda-2\lambda_1) \beta_3+\rho^2\!\sin x\beta_4.
\eea
The third set, shown on the right in Fig.~\ref{fig:c21bc2b}, is the same as the second, but the colours have been interchanged. This leads to the following discrete contour integral
\be
 \sum_{i}F(z_i)\Delta z_i
  = q_1\overline{\zeta}\beta_1-q_1\zeta\beta_2+q_1\overline{q}_1\overline{\zeta}\beta_3 - n_2 q\zeta\beta_4,
\label{eqn:c21bceqn3}
\ee
whose real and imaginary parts are
\bea
\!\!\!\!R_3 &\!\!\!\!=\!\!\!\!& \rho\cos(x+4\lambda+2\lambda_1)\beta_1-\rho\cos(x-4\lambda-2\lambda_1)\beta_2+\rho^2\cos x \beta_3 -n_2\rho\cos(x-4\lambda-2\lambda_1)\beta_4,  \\[.1cm]
\!\!\!\!I_3 &\!\!\!\!=\!\!\!\!& \rho\sin(x+4\lambda+2\lambda_1)\beta_1+\rho\sin(x-4\lambda-2\lambda_1)\beta_2+\rho^2\sin x \beta_3+n_2\rho\sin(x-4\lambda-2\lambda_1)\beta_4.
\eea

In the final two sets of configurations, shown in Fig.~\ref{fig:c21bc2c}, the defect has passed through the boundary before entering the boundary plaquette, but the two loop segments form a closed loop that crosses itself and has total winding angle $0$. The resulting discrete contour integrals are
\be
  \sum_{i}F(z_i)\Delta z_i
  = -q_1\zeta\beta_1+q_1\overline{\zeta}\beta_2-q_1^2\zeta\beta_3+\overline{q}_1n_3\overline{\zeta}\beta_4
\ee
and 
\be
  \sum_{i}F(z_i)\Delta z_i
  = q_1\overline{\zeta}\beta_1-q_1\zeta\beta_2+n_3\overline{q}_1\overline{\zeta}\beta_3-q_1^2\zeta\beta_4.
\ee
Taking real and imaginary components of each of the above gives 
\bea
R_4 &=& -\rho\cos(x-4\lambda-2\lambda_1)\beta_1+\rho\cos(x+4\lambda+2\lambda_1)\beta_2-
\rho^2\cos(x-4\lambda-4\lambda_1)\beta_3\nonumber\\
&&\mbox{} +\rho\, n_3\cos(x-4\lambda-2\lambda_1)\beta_4, \\[.1cm]
I_4 &=& \rho\sin(x-4\lambda-2\lambda_1)\beta_1+\rho\sin(x+4\lambda+2\lambda_1)\beta_2+
\rho^2\sin(x-4\lambda-4\lambda_1)\beta_3\nonumber\\
&&\mbox{}+\rho\, n_3\sin(x-4\lambda-2\lambda_1)\beta_4
\eea
and
\bea
R_5 &=& \rho\cos(x+4\lambda+2\lambda_1)\beta_1-\rho\cos(x-4\lambda-2\lambda_1)\beta_2+
n_3\rho\cos(x-4\lambda-2\lambda_1)\beta_3\nonumber\\
&&\mbox{} -\rho^2 \cos(x-8\lambda-4\lambda_1)\beta_4, \\[.1cm]
I_5 &=& \rho\sin(x+4\lambda+2\lambda_1)\beta_1+\rho\sin(x-4\lambda-2\lambda_1)\beta_2+
n_3\rho\sin(x-4\lambda-2\lambda_1)\beta_3\nonumber\\
&&\mbox{}+\rho^2 \sin(x-8\lambda-4\lambda_1)\beta_4.
\eea
Taking only the real expressions or only the imaginary expressions and setting each to zero ($R_1=R_2=R_3=R_4=R_5=0$ or $I_1=I_2=I_3=I_4=I_5=0$), we obtain a linear system in the unknown Boltzmann weights. One readily observes that only $R_1$, $R_2$ and $R_3$ are linearly independent and, similarly, $I_1$, $I_2$ and $I_3$ are also linearly independent. These are, in fact, the only independent equations arising from the boundary.

Now, considering the real equations and solving for the weights, we thus find
\be
\begin{array}{rcl}
 \beta_1(x) &=& n_1\cos(x-4\lambda+4\lambda_1), \\[.1cm]
 \beta_2(x) &=& \beta_1(x), \\[.1cm]
 \beta_3(x) &=& 2\sin4\lambda_1\sin x, \\[.1cm]
 \beta_4(x) &=& \beta_3(x).
\end{array}
\ee
Alternatively, setting the imaginary components to zero yields 
\be
\begin{array}{rcl}
 \beta_1(x) &=& n_1\sin(x-4\lambda+4\lambda_1), \\[.1cm]
 \beta_2(x) &=& - \beta_1(x), \\[.1cm]
 \beta_3(x) &=& -2\sin4\lambda_1\cos x, \\[.1cm]
 \beta_4(x) &=& -\beta_3(x).
\end{array}
\ee
We have verified that both of these sets of solutions solve the corresponding reflection equation.

\section{Conclusion}
\label{SecConcl}

We have shown how to obtain integrable boundary weights in the dilute $O(n)$ and $C_2^{(1)}$ loop models by imposing simple boundary conditions on a discretely holomorphic parafermionic observable. Interestingly, in each model, two sets of solutions to the corresponding reflection equation are obtained, according to whether we require the real or imaginary part of the discrete contour integral of the observable along the boundary to vanish.

In the case of non-diagonal boundaries, there are several types of loops, and we are forced to define new discretely holomorphic observables that incorporate additional parameters associated with boundary interactions. These parameters are fixed by requiring that (i) the discrete holomorphicity property in the bulk be preserved, and (ii) that a parameter which we interpret as the weight of a new type of loop be real.

As is the case in the bulk, the discrete holomorphicity approach involves solving a simple linear system of equations and thus bypasses the complicated non-linear functional equations arising from the reflection equation.  Once the solutions are obtained, it is straightforward to verify that they do indeed satisfy the reflection equation. This approach has allowed us to find, in a very direct way, new integrable boundary weights for both the dilute $O(n)$ and $C_2^{(1)}$ loop models.

A natural extension of this work is to define discretely holomorphic observables in other geometries such as the infinite strip and the cylinder. In the case of the cylindrical geometry, non-contractible loops appear, while in the infinite strip there are loops that touch both boundaries. 

We have only considered \textit{properties} of discretely holomorphic observables, but it is also of interest whether expectation values of these observables can be evaluated in closed form. This was, for example, achieved in~\cite{GierNP} for a particular observable in the Temperley-Lieb loop model on an inhomogeneous lattice and with open boundaries. 

Finally, a number of authors~\cite{DuminilSmir, Beatona, Beatonb} have found discretely holomorphic observables to be an important tool in proving the location of the critical points for self-avoiding walks. In the present work, focus has been on the integrable points of the models. However, since the integrable and critical points are expected to coincide in many cases, the observables defined in this paper could also be of use in rigorously proving the location of the models' critical points. In~\cite{Elvey-Price}, an off-critical extension was developed to explore what can be said rigorously about critical exponents.

\section*{Acknowledgment}
We are grateful for financial support from the Australian Research Council. JR is supported under the Future Fellowship scheme, project number FT100100774. AL is supported by an Australian Postgraduate Award. Part of this work was completed during the visit of JdeG and AL to the US Mathematical Sciences Research Institute (MSRI, USA) during the Spring 2012 Random Spatial Processes Program. We warmly thank Imam ul Alam, Nick Beaton, Denis Bernard, Paul Fendley, Tony Guttmann, Jesper Jacobsen, Bernard Nienhuis and Paul Zinn-Justin for useful discussions and inspiration.

\appendix
\section{Reflection equations}
\label{AppReflection}

\begin{figure}
\begin{center}
\includegraphics[scale=0.35]{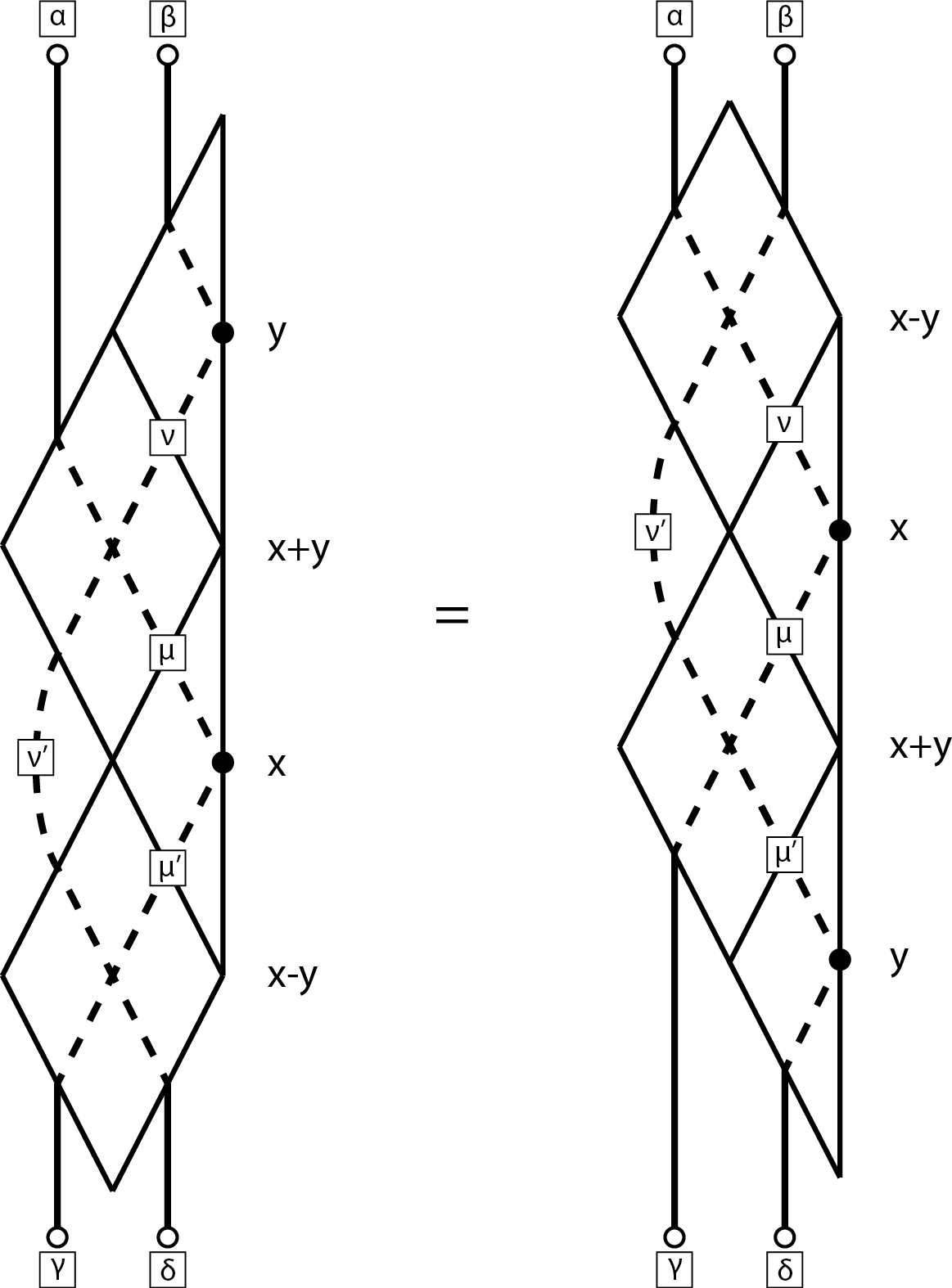}
\caption{A diagrammatic representation of the reflection equation. The spectral parameters associated with the plaquettes are shown. A functional equation in the Boltzmann weights is obtained for each choice of connectivity of the terminals, indicated here by open circles. }
\label{fig:byb}
\end{center}
\end{figure}

The Boltzmann weights of the bulk plaquettes are said to be $\textit{integrable}$ if they satisfy the Yang-Baxter equation. Similarly, the Boltzmann weights of the boundary plaquettes are integrable if they satisfy the \textit{reflection equation} of the form
\be
\sum_{\mu, \mu', \nu, \nu'}\!\!\mathrm{W}^{\beta}_{\nu} (x)\mathrm{W}^{\alpha\nu}_{\nu '\mu} (x+y)
 \mathrm{W}^{\mu}_{\mu '} (y) \mathrm{W}^{\nu '\mu '}_{\gamma\delta}(y-x) = \!\!\sum_{\mu, \mu', \nu, \nu'}\!\!\mathrm{W}^{\alpha\beta}_{\nu '\nu} (y-x)\mathrm{W}^{\nu}_{\mu} (y)\mathrm{W}^{\nu ' \mu}_{\gamma \mu '} (x+y) \mathrm{W}^{\mu '}_{\delta}(x).
\label{eqn:refeqn}
\ee
For our purposes, it is convenient to work with a diagrammatic representation of this equation, as shown in Fig.~\ref{fig:byb}. The indices $\alpha$, $\beta$, $\gamma$ and $\delta$ in (\ref{eqn:refeqn}) indicate the external connectivity, and we refer to them as the \textit{terminals}. The other indices, which occur once on each side of the equation, are summed over. The weights are functions of a spectral parameter. 

To obtain an equation in the Boltzmann weights, one first specifies an external connectivity, and subsequently sums over all choices of Boltzmann weights such that the connectivity between adjacent plaquettes matches. Each term in the ensuing equation thus arises from a choice of bulk and boundary plaquettes that are consistent with the given external connectivity. 

\begin{figure}
\begin{center}
\includegraphics[scale=0.4]{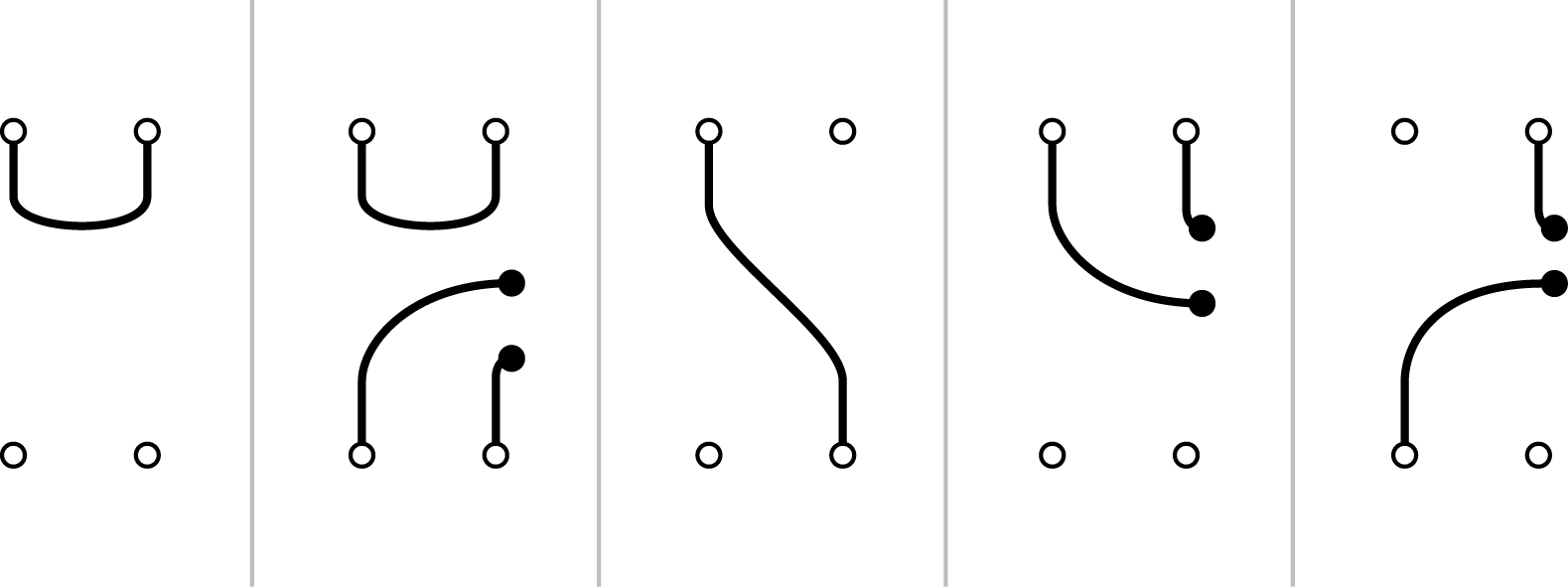}
\end{center}
\caption{The five connectivities leading to non-trivial functional equations for the dilute $O(n)$ model. The terminals are labelled by open circles. A black disk indicates that the associated terminal is connected to the boundary.}
\label{fig:ONbybcon}
\end{figure}

\begin{figure}
\begin{center}
\includegraphics[scale=0.4]{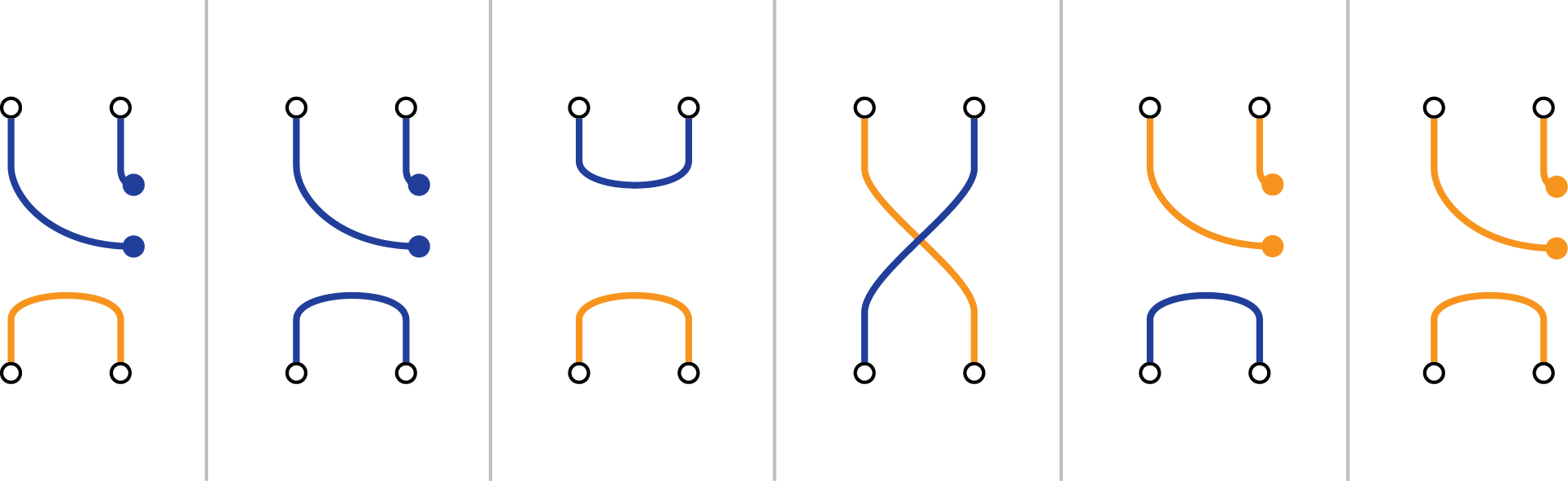}
\end{center}
\caption{The six connectivities of the terminals leading to non-trivial functional equations from the reflection equation for the $C_2^{(1)}$ model. A blue (orange) disk indicates that the corresponding terminal is connected to a $\beta_4$ ($\beta_3$) boundary plaquette.}
\label{fig:C21bybcon}
\end{figure}

From symmetry considerations, the only non-trivial equations arise from choices of external connectivity that are not invariant under interchanging the top and bottom terminals. For the $O(n)$ loop model, there are thus five non-trivial functional equations arising from the reflection equation. The connectivities of the diagrams are shown in Fig.~\ref{fig:ONbybcon}. As an example, Fig.~\ref{fig:ONbyb1} shows the set of diagrams giving rise to
\begin{figure}[ht]
\centering
\begin{picture}(300, 200)
\put(0,0){\includegraphics[scale=0.5]{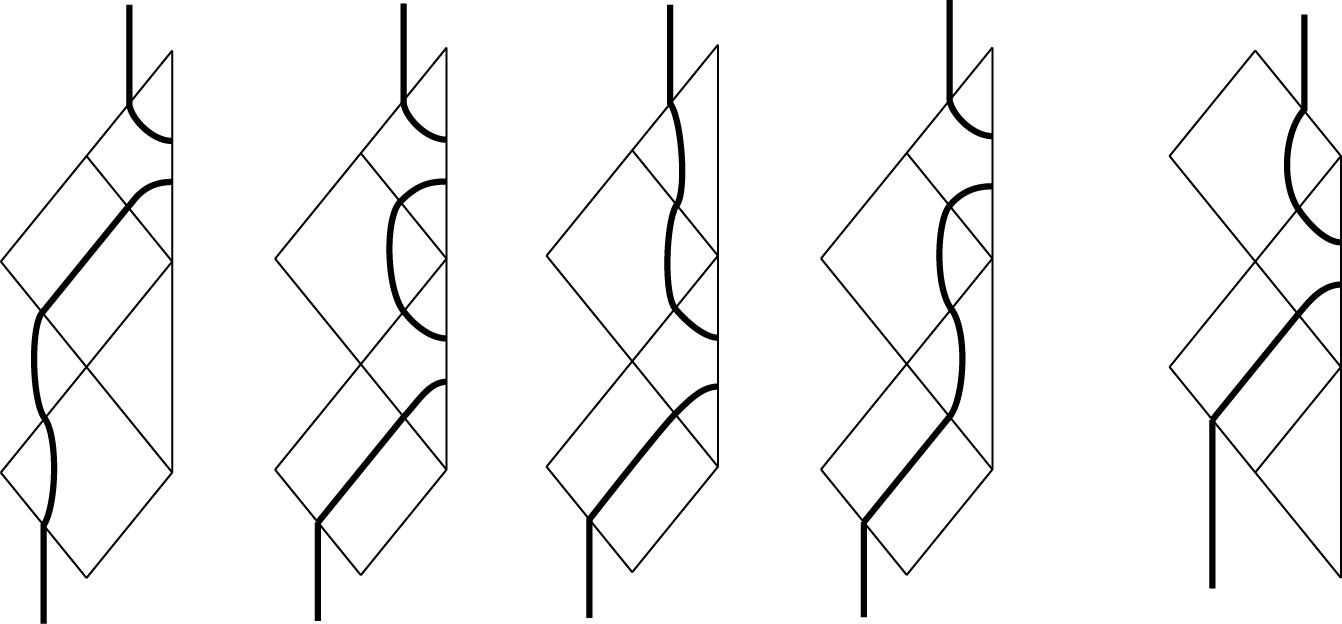}}
\put(50, 70){$+$}
\put(120, 70){$+$}
\put(190, 70){$+$}
\put(260, 70){$=$}
\end{picture}
\caption{The diagrams giving rise to (\ref{eqn:ONbyb}).}
\label{fig:ONbyb1}
\end{figure}
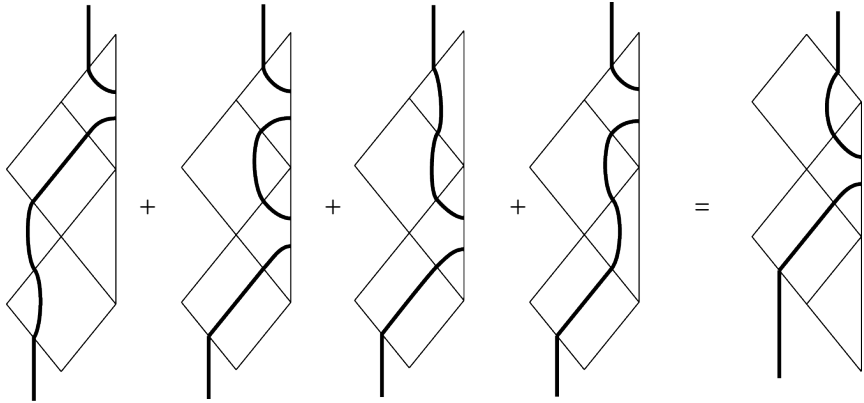
\begin{eqnarray}
&& \beta_3(y)v(x+y)\beta_1(x)u_1(x-y) + n_2\beta_3(y)u_1(x+y)\beta_3(x)v(x-y) \nonumber \\
&+& \beta_2(y)u_1(x+y)\beta_3(x)v(x-y) + \beta_3(y)u_1(x+y)\beta_2(x)v(x-y) \nonumber \\
&=& u_1(x-y)\beta_3(x)v(x+y)\beta_1(y).
\label{eqn:ONbyb}
\end{eqnarray}
For the $C_2^{(1)}$ loop model, there are six non-trivial functional equations to consider, without assuming that the weights are symmetric under interchanging colours. The connectivities of the terminals are shown in Fig.~\ref{fig:C21bybcon}.

\subsection{Generalised dilute \boldmath{$O(n)$} model}
\label{AppAsymmOn}

%
\begin{figure}
\begin{center}
\includegraphics[scale=0.3]{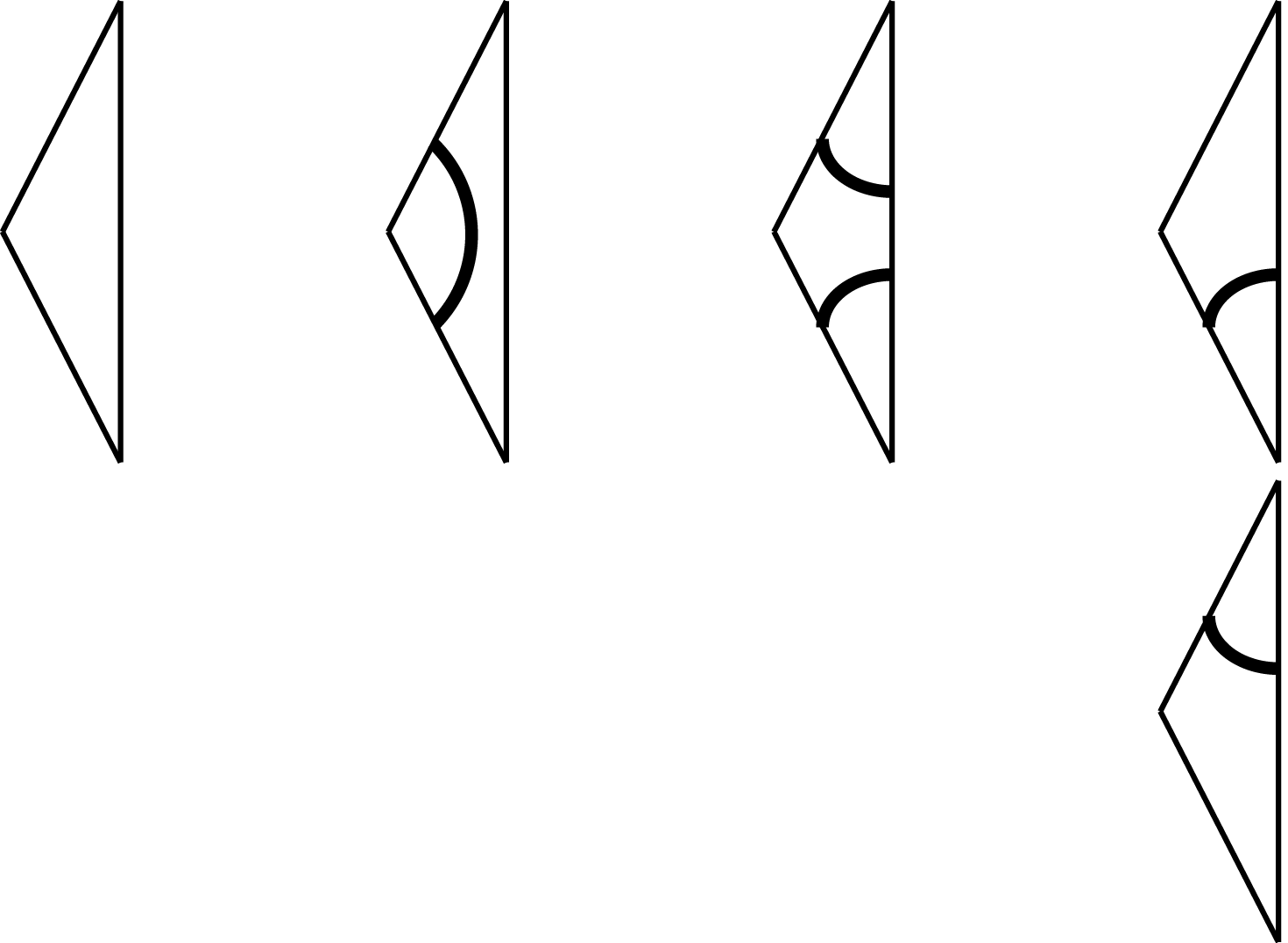}
\end{center}
\caption{The four types of boundary plaquettes in the generalised dilute $O(n)$ loop model. The corresponding plaquette weights are denoted by $\beta_1$, $\beta_2$, $\beta_3$ and $\beta_4$, respectively.}
\label{fig:ONabcd}
\end{figure}
One can extend the dilute $O(n)$ loop model described in Section~\ref{SecDilute} by including \textit{asymmetric} boundary conditions, as indicated to the right in Fig.~\ref{fig:ONabcd}. In keeping with the notation for the other boundary plaquettes, we denote the weight of an asymmetric boundary plaquette by $\beta_4$. A crucial consequence of the introduction of these new plaquettes is the possibility of forming boundary loops between two upper edges or between two lower edges of the boundary plaquettes. As we did for the parafermionic observable, we assign such loops the fugacity $n_3$. It is stressed that a defect need not be present for such loops to appear in the generalised model discussed here.

Assuming $\beta_4\neq0$, we find that there is no solution to the corresponding reflection equation unless
\be
 n_1=n_2=n_3.
\ee
In this case, 
\be
\begin{array}{rcl}
 \beta_1(x)&=&2\cos\lambda\sin(\tfrac{3}{2}\lambda+x) -k^2 n_1\sin(\tfrac{1}{2}\lambda+x)
   \sin(\tfrac{1}{2}\lambda-x)\sin(\tfrac{3}{2}\lambda-x),  \\[.1cm] 
 \beta_2(x)&=&\sin(\tfrac{3}{2}\lambda-x)\big[2\cos\lambda-k^2 n_1\sin^2(\tfrac{1}{2}\lambda-x)\big],
    \\[.1cm]
 \beta_3(x)&=&-k^2\sin2\lambda\sin2x\sin(\tfrac{1}{2}\lambda-x),  \\[.1cm]
 \beta_4(x)&=&k\sin2\lambda\sin2x,
\end{array}
\label{beta1234}
\ee
provides a one-parameter family of solutions, labelled by $k$. It follows, in particular, that there is no solution for which $\beta_4\neq0$ but $\beta_3=0$.

We observe that, for $k=0$, the solution (\ref{beta1234}) reduces to the solution (\ref{eqn:ondbc}). We also observe that the solution approaches the solution (\ref{betannn}) (with $n_1=n_2$) in the limit 
\be
 \lim_{k\to\infty}\frac{-\beta(x)}{k^2\sin(\tfrac{1}{2}\lambda-x)},\qquad
  \beta\in\{\beta_1,\beta_2,\beta_3\};\qquad
  \lim_{k\to\infty}\frac{-\beta_4(x)}{k^2\sin(\tfrac{1}{2}\lambda-x)}=0.
\ee


%

\begin{thebibliography}{99}
%
\bibitem{Smirnova} S.~Smirnov, \textit{Discrete complex analysis and probability}, Proc. Int. Congr. Math. (Hyderabad, India, 2010), 595--621, arXiv:1009.6077.
%
\bibitem{Smirnovb} S.~Smirnov, \textit{Towards conformal invariance of 2D lattice models}, Proc. Int.
Congr. Math. (Madrid, Spain, 2006), vol. II, 1421--1451, arXiv:0708.0032.
%
\bibitem{Duminil-CopinS} H.~Duminil-Copin and S.~Smirnov, \textit{Conformal invariance of lattice models}, arXiv:1109.1549.
%
\bibitem{BernardF} D.~Bernard and G.~Felder, \textit{Quantum group symmetries in two-dimensional lattice quantum field theory}, Nucl. Phys. \textbf{B365} (1991) 98--120.
%
\bibitem{SmirnovFermions} S.~Smirnov, \textit{Conformal invariance in random cluster models. I. Holomorphic fermions in the Ising model}, Ann. Math. \textbf{172} (2010) 1435--1467, arXiv:0708.0039.
%
\bibitem{DuminilSmir} H.~Duminil-Copin and S.~Smirnov, \textit{The connective constant of the honeycomb lattice equals $\sqrt{2+\sqrt{2}}$}, Ann. Math. \textbf{175} (2012) 1653--1665, arXiv:1007.0575.
%
\bibitem{Nienhuis} B.~Nienhuis, \textit{Exact critical point and critical exponents of $O(n)$ models in two dimensions}, Phys. Rev. Lett. \textbf{49} (1982) 1062--1066.
%
\bibitem{RivaCardy} V.~Riva and J.~Cardy, \textit{Holomorphic parafermions in the Potts model and SLE}, J. Stat. Mech. 0612 (2006) P12001, arXiv:cond-mat/0608496.
%
\bibitem{LottiniR} S.~Lottini and M.A.~Rajabpour, \textit{Ashkin-Teller model on the iso-radial graphs}, J. Stat. Mech. 2010 P06027, arXiv:1003.6080.
%
\bibitem{IkhlefR} Y.~Ikhlef and M.A.~Rajabpour, \textit{Discrete holomorphic parafermions in the Ashkin-Teller model and SLE}, J. Phys. A: Math. Theor. \textbf{44} (2011) 042001, arXiv:1009.3374.
%
\bibitem{RajabCardy} M.A.~Rajabpour and J.~Cardy, \textit{Discretely holomorphic parafermions in lattice $Z_N$ models}, J. Phys. A: Math. Theor. \textbf{40} (2008) 14703--14713, arXiv:0708.3772.
%
\bibitem{IkhlefCardy} Y.~Ikhlef and J.~Cardy, \textit{Discretely holomorphic parafermions and integrable loop models}, J. Phys. A: Math. Theor. \textbf{42} (2009) 102001, arXiv:0810.5037.
%
\bibitem{Cardy} J.~Cardy, \textit{Discrete holomorphicity at two-dimensional critical points}, J. Stat. Phys. \textbf{137} (2009) 814--824, arXiv:0907.4070.
%
\bibitem{Tanhayi-AhariR} M.~Tanhayi-Ahari and S.~Rouhani, \textit{Discrete holomorphic parafermions in the eight vertex model}, arXiv:1209.4253.
%
\bibitem{AlamBatch} I.T.~ul~Alam and M.T.~Batchelor, \textit{Integrability as a consequence of discrete holomorphicity: the $Z_N$ model}, arXiv:1207.3883.
%
\bibitem{Nienhuis90} B.~Nienhuis, \textit{Critical spin-1 vertex models and $O(n)$ models}, Int. J. Mod. Phys. B \textbf{4} (1990) 929--942.
%
\bibitem{Nienhuis2} B.~Nienhuis, \textit{Critical and multicritical $O(n)$ models}, Physica A \textbf{163} (1990) 152--157.
%
\bibitem{Sklyanin} E.K.~Sklyanin, \textit{Boundary conditions for integrable quantum systems}, J. Phys. A: Math. Gen. \textbf{21} (1988) 2375--2389. 
%
\bibitem{BPO96} R.E.~Behrend, P.A.~Pearce and D.L.~O'Brien, \textit{Interaction-round-a-face models with fixed boundary conditions: the ABF fusion hierarchy}, J. Stat. Phys. \textbf{84} (1996) 1--48, arXiv:hep-th/9507118.
%
\bibitem{PRZ06} P.A.~Pearce, J.~Rasmussen and J.-B.~Zuber, \textit{Logarithmic minimal models}, J. Stat. Mech. 0611 (2006) P11017, arXiv:hep-th/0607232.
%
\bibitem{PRR08} P.A.~Pearce, J.~Rasmussen and P.~Ruelle, \textit{Integrable boundary conditions and $\mathcal{W}$-extended fusion in the logarithmic minimal models $\mathcal{LM}(1,p)$}, J. Phys. A: Math. Theor. \textbf{41} (2008) 295201, arXiv:0803.0785.
%
\bibitem{Beatona} N.~Beaton, M.~Bousquet-M\'elou, J.~de~Gier, H.~Duminil-Copin and A.J.~Guttmann, \textit{The critical fugacity for surface adsorption of SAW on the honeycomb lattice is $1+\sqrt{2}$}, arXiv:1109.0358.
%
\bibitem{Batchelor} M.T.~Batchelor and C.M.~Yung, \textit{Exact results for the adsorption of a flexible self-avoiding polymer chain in two dimensions}, Phys. Rev. Lett. \textbf{74} (1995) 2026--2030, arXiv:cond-mat/9410082.
%
\bibitem{Beatonb} N.~Beaton, \textit{The critical surface fugacity of self-avoiding walks on a rotated honeycomb lattice}, arXiv:1210.0274. 
%
\bibitem{Ikhlef} Y.~Ikhlef, \textit{Discretely holomorphic parafermions and integrable boundary conditions}, J. Phys. A: Math. Theor. {\bf 45} (2012) 265001, arXiv:1202.6265.
%
\bibitem{DJSa} J.~Dubail, J.L.~Jacobsen and H.~Saleur, \textit{Conformal two-boundary loop model on the annulus}, Nucl. Phys. \textbf{B813} (2009) 430--459, arXiv:0812.2746.
%
\bibitem{DJSb} J.~Dubail, J.L.~Jacobsen and H.~Saleur, \textit{Conformal boundary conditions in the critical $O(n)$ model and dilute loop models}, Nucl. Phys. \textbf{B827} (2010) 457--502, arXiv:0905.1382.
%
\bibitem{WarnaarNienhuis} S.O. Warnaar and B. Nienhuis, \textit{Solvable lattice models labelled by Dynkin diagrams}, J. Phys. A: Math. Gen. \textbf{26} (1993) 2301--2316, arXiv:hep-th/9301026.
%
\bibitem{GierNP} J. de Gier, B. Nienhuis and A. Ponsaing, \textit{Exact spin quantum Hall current between boundaries of a lattice strip}, Nucl. Phys. \textbf{B838} (2010) 371--390, arXiv:1004.4037.
%
\bibitem{Elvey-Price} A.~Elvey~Price, J.~de~Gier, A.J.~Guttmann and A.~Lee, \textit{Off-critical parafermions and the winding angle distribution of the $O(n)$ model}, J. Phys. A: Math. Theor. \textbf{45} (2012) 275002, arXiv:1203.2959.
%
\end{thebibliography}
\end{document}